\documentclass[twocolumn,trackchanges]{aastex7}

\begin{document}

\title{A Common Synchrotron Origin for Prompt Gamma-Ray and Soft X-Ray Emission in GRBs: Evidence from Joint Spectral Analysis}

\correspondingauthor{He Gao}
\email{gaohe@bnu.edu.cn}

\author[orcid=0009-0006-9824-2498,gname=Ziming,sname=Wang]{Ziming Wang}
\affiliation{School of Physics and Astronomy, Beijing Normal University, Beijing 100875, China}
\affiliation{Institute for Frontier in Astronomy and Astrophysics, Beijing Normal University, Beijing 102206, China}
\email{zmwang489@mail.bnu.edu.cn}

\author[orcid=0009-0007-9068-2752,gname=Chenyu,sname=Wang]{Chenyu Wang}
\affiliation{Department of Astronomy, Tsinghua University}
\email{wang-cy22@mails.tsinghua.edu.cn}

\author[orcid=0000-0003-2516-6288, gname= He, sname='Gao']{He Gao}
\affiliation{School of Physics and Astronomy, Beijing Normal University, Beijing 100875, China}
\affiliation{Institute for Frontier in Astronomy and Astrophysics, Beijing Normal University, Beijing 102206, China}
\email[show]{gaohe@bnu.edu.cn}

\author[orcid=0000-0001-7584-6236, gname=Hua,sname=Feng]{Hua Feng}
\affiliation{State Key Laboratory of Particle Astrophysics, Institute of High Energy Physics, Chinese Academy of Sciences, 100049, Beijing, China}
\email{hfeng@ihep.ac.cn}

\author[orcid=0000-0002-0823-4317,gname=An,sname=Li]{An Li}
\affiliation{School of Physics and Astronomy, Beijing Normal University, Beijing 100875, China}
\affiliation{Institute for Frontier in Astronomy and Astrophysics, Beijing Normal University, Beijing 102206, China}
\email{202221160007@mail.bnu.edu.cn}

\author[orcid=0000-0002-0633-5325,gname=Lin,sname=Lin]{Lin Lin}
\affiliation{School of Physics and Astronomy, Beijing Normal University, Beijing 100875, China}
\affiliation{Institute for Frontier in Astronomy and Astrophysics, Beijing Normal University, Beijing 102206, China}
\email{llin@bnu.edu.cn}

\author[orcid=0009-0000-6275-3452, gname= Song-Yu, sname='Shen']{Song-Yu Shen}
\affiliation{School of Physics and Astronomy, Beijing Normal University, Beijing 100875, China}
\affiliation{Institute for Frontier in Astronomy and Astrophysics, Beijing Normal University, Beijing 102206, China}
\email{syshen@mail.bnu.edu.cn}  








\begin{abstract}

The recent launches of the Einstein Probe (EP) and the Space Variable Objects Monitor (SVOM) mission have led to the detection of a growing number of long GRBs with significant, early soft X-ray flux during their gamma-ray emission, prompting the question of whether their multi-band prompt emission shares a common origin in region and mechanism. To address this, we utilize the 20-year Swift archival data, which provides a substantial sample of joint soft X-ray and gamma-ray observations, enabling a systematic joint spectral study. We resolve 110 temporal pulses from 46 GRBs and find that a single power-law model with a low-energy break or cutoff adequately describes the prompt spectra from 150 keV down to 0.5 keV. More than half of the sample pulses require a break around a few keV, with average spectral indices $\langle \alpha_1 \rangle = -0.88$ and $\langle \alpha_2 \rangle = -1.46$ consistent with synchrotron radiation in a marginally fast-cooling regime. The observed spectral evolution and the distribution of indices support a single-emission-region origin, where the varying spectral shapes are largely governed by the evolution of the synchrotron cooling frequency $\nu_c$ and the effect of finite emission width. The observed differences in the temporal behavior between X-ray and gamma-ray light curves can be naturally explained by this spectral evolution across the broad band.

\end{abstract}



\section{Introduction} 
Gamma-Ray Bursts (GRBs) are transient, luminous burst phenomenon of gamma-ray radiation originates at cosmological distances. 
Observationally, GRB emission are divided into prompt and afterglow phase, where prompt is first recognized as the temporal phase during which excessive sub-MeV emission is detected \citep{2018pgrb.book.....Z}. 
`X-ray flashes' (XRFs; \citep{2003A&A...400.1021B,2005ApJ...629..311S}) are GRB-like events whose emission is mostly in X-rays, while X-ray–rich GRBs (XRRs; \citep{2001grba.conf...16H}) represent an intermediate class in between. 
Studies on these events have suggested a same origin as GRBs for these phenomenon \citep{2006A&A...460..653D}, expanding GRB energies toward lower-energy end. 
The prompt emission is now more broadly defined to include multi-wavelength emissions that share a common origin with gamma rays observed during the same period \citep{2005Natur.435..178V}.
The long burst GRB 240315A/EP240315a, recently detected by Konus-Wind \citep{1995SSRv...71..265A}, Swift/BAT \citep{2005SSRv..120..143B}, and the Einstein Probe \citep{2022hxga.book...86Y}, showed a soft X-ray component with an earlier onset, longer duration, and more complex structure than the gamma-rays \citep{2025NatAs...9..564L}. 
The subsequent detection of similar bursts motivate a deeper understanding of prompt emission origins in the soft X-ray band.

%
%

%
Multiple radiative process that originated from different spatial regions may contribute during the GRB prompt phase, including quasi-thermal photosphere radiation \citep{Rees2005ApJ...628..847R,Ryde2009ApJ...702.1211R}, synchrotron from accelerated electrons \citep{1994ApJ...430L..93R,1996ApJ...473..204S} as well as inverse Compton scattering \citep{2019Natur.575..459M}. 
The typical prompt spectra show Band function \citep{1993ApJ...413..281B} power-law slopes of $\alpha \sim -1$ and $\beta \sim -2.5$, with a broad peak energy range from few hundred keV to few MeV, confirmed by multiple missions such as BATSE \citep{2000ApJS..126...19P,2006ApJS..166..298K} and \textit{Fermi}-GBM \citep{2011A&A...530A..21N,2021ApJ...913...60P}.
Some GRB spectra can be well described by the synchrotron radiation model \citep{2024zhenyuSYN}.
However, the average spectral index below the peak energy does not align with the predictions of standard synchrotron value \citep{2019Ravasio}. 
This discrepancy was attributed to either the presence of a photosphere component within the observed energy band \citep{Preece_1996,Meszaros:2002vh,Peng2014ApJ...795..155P,2006ApJ...642..995P,2025arXiv250104906W} or modifications to the synchrotron emission process itself \citep{2011A&A...526A.110D}.

The \textit{Neil Gehrels \textit{Swift} Observatory} \citep{2004ApJ...611.1005G,Gehrels2009ARA&A..47..567G} equipped with the Burst Alert Telescope (BAT; \cite{2005SSRv..120..143B}) and the X-ray Telescope (XRT; \cite{2005SSRv..120..165B}), along with its fast slewing capability, has, over its 20 years of operation, accumulated many cases with early-time X-ray observations during prompt phase \citep{Vaughan_2006,Morris_2007,Cusumano_2006,10.1111/j.1365-2966.2007.12763.x}. 
Few works utilizing \textit{Swift} and \textit{Fermi} data \citep{Zheng2012ApJ...751...90Z,Gendre2012ApJ...748...59G,Oganesyan2017ApJ...846..137O,2018A&A...616A.138O,Ravasio_2019} highlight that adding an extra low-energy break at around 10keV would fit observed spectra well.
Their results also suggest synchrotron radiation as origin of GRB emission. 
Combinations and varied significance of radiative mechanisms are influenced by various physical parameters of the GRB, such as jet composition, magnetic field strength, and energy dissipation processes \citep{ZHANG_2014}, which would help us probe the properties of the GRB origin. 
Delving into low-energy bands to verify the radiative models would surely benefit theoretical studies.

In this work, we aim to utilize the most up-to-date sample of \textit{Swift} GRBs (as of March 2025) to joint fit the resolved prompt spectra from 0.5keV to 150keV. 
We selected 46 GRB with a more relaxed criteria compared to preexisting works which includes later and fainter emission epochs into the sample. 
We find that, in most of our spectral intervals, X-ray data does not fit the extrapolation of gamma-ray data. 
We fitted popular power-law models and a thermal component to make comparison. 
Our results prefer the same BPL model containing a low-energy break, as mentioned by above previous works. 
Fitted power-law segment below the breaks are slightly softer than previous results.
We also confirm a spectral softening with increasing normalized time within the GRB $T_{90}$ interval. 
The result we found fits quite well within the properties and ranges of a single-component fast-cooling synchrotron emission. 

The structure of this paper is as follows.
In Section \ref{sample-selection}, we describe the selection method of our sample of pulses. Section \ref{spectral-analysis} presents the methods and process of data extraction and spectral analysis, followed by results of the spectral fit presented in Section \ref{results}. 
We discuss our findings and summarize this work in Section \ref{summary}. 

\section{Sample Selection and temporal binning} \label{sample-selection}
GRB prompt emission light curves can be decomposed into multiple overlapping pulses, which may show different spectral properties while observed in different energy ranges \citep{2018pgrb.book.....Z}. 
Our objective is to investigate the spectral features of GRB prompt emission observed simultaneously by the two instruments onboard \textit{Swift}. 
We first obtain a list of all GRBs up to March 2025 from \textit{\textit{Swift} GRB Table}\footnote{\url{https://Swift.gsfc.nasa.gov/archive/grb_table/}}. 
241 GRBs were selected with overlapping observation periods from both XRT and BAT.
To identify temporal pulses in GRBs, we build normalized 1-second count-rate (see 3.1) curves.
5-$\sigma$ binned flux curves\footnote{obtained from \url{https://www.Swift.ac.uk/burst_analyser/}} are also used to inspect emission periods. 
We perform a Bayesian block algorithm \citep{1998ApJ...504..405S} on the count-rate curve to obtain a series of edges. 
To focus on the identical temporal pulses with both `rise and decay' slopes (hereafter referred to as pulses), we define resolved spectral interval start and stop time as rate minimum locations of a pulse, with reference to the Bayesian Block edges. 
We impose a count rate signal-to-noise ratio (SNR) threshold of 5 for all pulses for reliable spectral analysis. 
For pulse intervals that failed to meet this criterion, we either discard or merge them with adjacent pulse intervals. 
GRBs without a single qualified pulse were excluded from our sample. 
By steps above, we conclude a final sample of 46 GRBs with at least one significant pulse observed in both bands\footnote{Slew stage of XRT data is not considered. All data are observed in XRT windowed timing(WT) mode. }. 

A total of 113 pulses were obtained from our temporal binning criteria. 
We build products in both bands and perform subsequent spectral analysis within these pulses.
Quiescent times and some insignificant intervals (i.e. an isolated pulse of SNR$<5\sigma$, intervals containing only decaying slope) are ruled out by our selection. 
We show binned-by-pulse light curves in Appendix \ref{all-lc}.
%


\section{Data Reduction and Spectral Analysis} \label{spectral-analysis}
Data products are build from level 2 event files and auxiliary files \citep{2009MNRAS.397.1177E}, obtained from the \textit{HEASARC} FTP site\footnote{\url{https://heasarc.gsfc.nasa.gov/FTP/swift/data/obs/}} for the selected GRBs. Data has been processed using \texttt{HEASoft}\footnote{\citep{2014ascl.soft08004N}} package (V6.33.1). The latest release of calibration files to date were adopted for BAT(\texttt{CALDB 2023-06-07}) and XRT(\texttt{CALDB 2024-05-22}). 

\subsection{\textit{Swift} Data Reduction}
Spectrum and 1-second binned count-rate light curve for all intervals in BAT 15-150keV were built using \texttt{batbinevt} and applied correction following the standard analysis threads provided by \textit{GSFC}\footnote{\url{https://Swift.gsfc.nasa.gov/analysis/threads/bat_threads.html}}. An individual response matrix was created for each pulse interval. 
The same data products in XRT 0.4-10keV were built using \texttt{xrtproducts} following the \textit{Swift} XRT Data Reduction Guide (Version 1.2 April 2005)\footnote{M. Capalbi et al. \url{https://Swift.gsfc.nasa.gov/analysis/xrt_swguide_v1_2.pdf}} as well as \textit{SDC} XRT Analysis Threads\footnote{\url{https://www.Swift.ac.uk/analysis/xrt/}}. Only cleaned event files in WT pointing mode (\textit{*xwtw2po\_cl.evt}) were used. A specific set of exposure files were built by \texttt{xrtexpomap} for each interval in case of source slightly moving on the detector plain among multiple exposures. 

\begin{table}
\caption{Count-rate dependent source exclusion region radius for windowed timing mode pile-up elimination}
\label{tab:pileup}
\begin{tabular}{|c|c|} \hline
Peak rate $R_{\mathrm{wt}}$ (count s$^{-1}$) & Exclusion radius \\ 
 & pixels(arcsec) \\ \hline
$R_{\mathrm{wt}} < 100$ & 0 \\ \hline
$100 < R_{\mathrm{wt}} < 200$ & 1 (2.38$^{\prime\prime}$) \\ \hline
$200 < R_{\mathrm{wt}} < 400$ & 2 (4.75$^{\prime\prime}$) \\ \hline
$400 < R_{\mathrm{wt}} < 600$ & 4 (9.50$^{\prime\prime}$) \\ \hline
$600 < R_{\mathrm{wt}} < 1000$ & 8 (19.00$^{\prime\prime}$) \\ \hline
$R_{\mathrm{wt}} > 1000$ & 10 (23.75$^{\prime\prime}$) \\ \hline
\end{tabular}
\end{table}

\begin{figure}[ht!]
    \centering
    \includegraphics[width=0.47\textwidth]{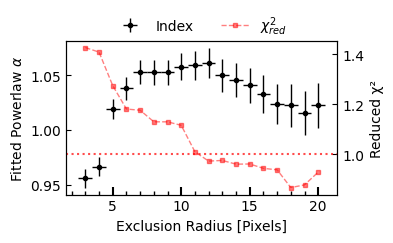}
    \caption{GRB 241030A: fitted single power-law indices of integrated x-ray spectrum, affected by the selection of different exclusion region radius. }
    \label{fig:depile}
\end{figure}

Source region was set to an annuli with an outer bound of 30 pixels and different inner bound according to pile-up status caused by high source count rate. 
Since every of our observation interval are within early prompt stage of the GRBs, the flux densities are at intense stage. Most of the count rate exceeds 150 photons $s^{-1}$ thus are heavily piled up. 
We adopted the method listed in \cite{2006A&A...456..917R} to mitigate pile-up by excluding piled-up source region. 
For each interval, a non-exclusion run of 30 central pixels is first performed to find the maximum count rate, then the size of the exclusion region is defined accordingly to make sure the peak rate does not exceed 150cts/s within the interval. 
The number of pixels excluded from the source center are set referring to values listed in \texttt{xrtgrblc}\footnote{\url{https://heasarc.gsfc.nasa.gov/docs/software/lheasoft/help/xrtgrblc.html}} help file, shown in Table \ref{tab:pileup} and Figure \ref{fig:depile}.

Background extraction region was set to an annuli with radius from 80 to 120 pixels. 
Background scaling was done based on the area of source and background extraction region. 
In order to adopt $\chi^2$ statistics, energy channels were grouped by \texttt{grppha} tool requiring a minimum of 20 photons per energy bin. 

\subsection{Spectral Analysis}
A python code was built using the PyXspec (2.1.4)\footnote{\url{https://heasarc.gsfc.nasa.gov/docs/xanadu/xspec/python/html/index.html}} package to perform batch spectral fitting. Energy ranges are set to 0.5-10keV for XRT, and 15-150keV for BAT. Intercaliberation uncertainties between the two instruments are within relatively low range, allowing us to use data products to perform joint spectral fitting without introducing a multiplying factor \citep{2006NCimB.121.1463C}. 

Observed GRB non-thermal spectra are usually well described by one or a few power-law segments. 
In this work, we test the most simple power-law (SPL) model described by a photon index $\alpha$, a cutoff power-law (CPL) with an exponential cutoff at $E_{\text{cut}}$, and a broken power-law (BPL) with two segments divided by an break at $E_{\text{break}}$. 
%
%
A subdominant thermal component was also statistically favored in some GRBs \citep{100507thermal,160107Athermal}. 
Scenarios of an additive xspec blackbody (BB) model are also tested based on above power-law models, giving us 6 models to consider in total: SPL, CPL, BPL, (BB)SPL, (BB)CPL, (BB)BPL. 
Gamma-ray observation typically lies in the energy range of $10^2 \sim 10^3$keV, corresponding to the majority of cutoff $E_{\text{peak}}$ and the `Band break' region at over 100keV. 
Regarding the highest reliable limit of BAT spectra at 150keV and limited photon counts in many pulses, we do not seek to well constrain $E_{\text{peak}}$ values in this work.
However some late-time GRB pulses have displayed a relatively low energy peak which were observed in BAT band. 

$\chi^{2}$ statistics are often used to reflect the wellness-of-fit with a reduced chi-squared ($\chi^{2}_{\rm red} = \chi^{2}/dof$). 
Fit with $\chi^{2}_{\rm red} > 2$ are considered as bad-fit and discarded. 
However, improving the fitting result by introducing extra parameters such as power-law segments, break energies, or a cutoff energy would increase the complicity of the model. 
In this work, we determine whether an extra spectral feature is crucial by adopting \textit{F-test} \citep{ftest} and \textit{BIC} test \citep{2007MNRAS.377L..74L} to penalize model complexity, thereby avoid overfitting. 
Our comparison procedure starts with SPL which is the simplest of the six models. 
The \textit{F-test} is used to compare between nested models with a rather strict 5$\sigma$ significance level requirement to select a more complex model. 
For example, if a CPL model improves the SPL fit by $>5\sigma$, CPL becomes the current best model. 
Between non-nested models such as (BB)SPL and BPL, we required $\Delta$BIC$>$10 to select a more complex model. 
The only models with comparable complexity are (BB)SPL and BPL ($k=4$); among them, BPL is considered first.
%
%

Absorption was taken account using the Tuebingen-Boulder ISM absorption model integrated in \texttt{XSPEC} (\cite{1996ASPC..101...17A}) function \textit{tbabs} and \textit{ztbabs} for galactic and intrinsic absorption. Element abundances are set to Wilms \citep{2000ApJ...542..914W}. 
Galactic nH column densities along the burst direction were well documented by \cite{2005A&A...440..775K}. We obtained nH values from the \textit{UKSSDC} online check-up tool\footnote{\url{https://www.swift.ac.uk/analysis/nhtot/}} and list them in Appendix \ref{absorption_params}. 
Intrinsic absorption is hard to estimate by fitting and will display dramatic variations itself if left as a free parameter, thus impact the spectral parameters. 
Also, the possibility of circumburst nH being affected by photonionization caused by the burst radiation process is suggested by \cite{2002MNRAS.330..383L}. 
Considering the poor fit constants and possible variation in its nature, we fix intrinsic absorption parameters (nH, redshift) to the values obtained by late-time XRT observation fitting given by SDC, where light curve and spectrum are usually well-described by a single power-law slope. 

We find 3 out of 113 batch processed results exceed physical model limits. GRB201026A (pulse1) and GRB241030A (pulse9) have power-law indices exceed typical range of $-3 < \alpha < 0 $. GRB220101A (pulse4) have a BPL fit showing steeper decay in $\gamma$-ray band than in X-ray band. These three pulses are dimmed bad fit and discarded from our sample. 

Resolved light curve and spectrum along with best-fit model for GRB 241030A is shown in Figure \ref{fig:241030lc},\ref{fig:241030spec}. Figures for all selected GRBs are shown in Appendix \ref{all-spec}. 
Fit results and parameters of all pulses are listed in Appendix \ref{all-params}. 

\begin{figure}
    \includegraphics[width=0.47\textwidth]{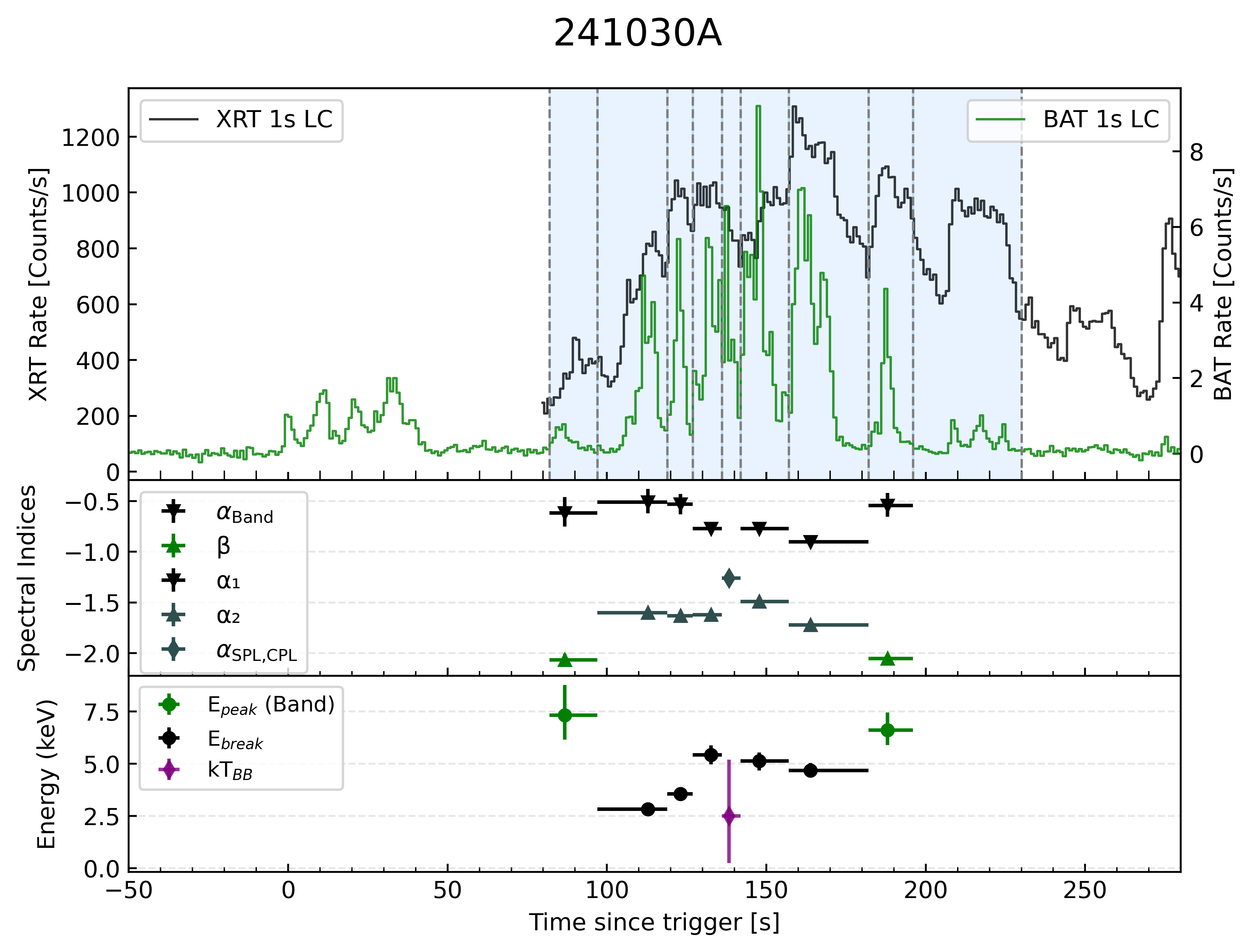}
    \caption{GRB 241030A: resolved light curve and best-fit spectral parameters}
    \label{fig:241030lc}
\end{figure}
\begin{figure}
    \includegraphics[width=0.47\textwidth]{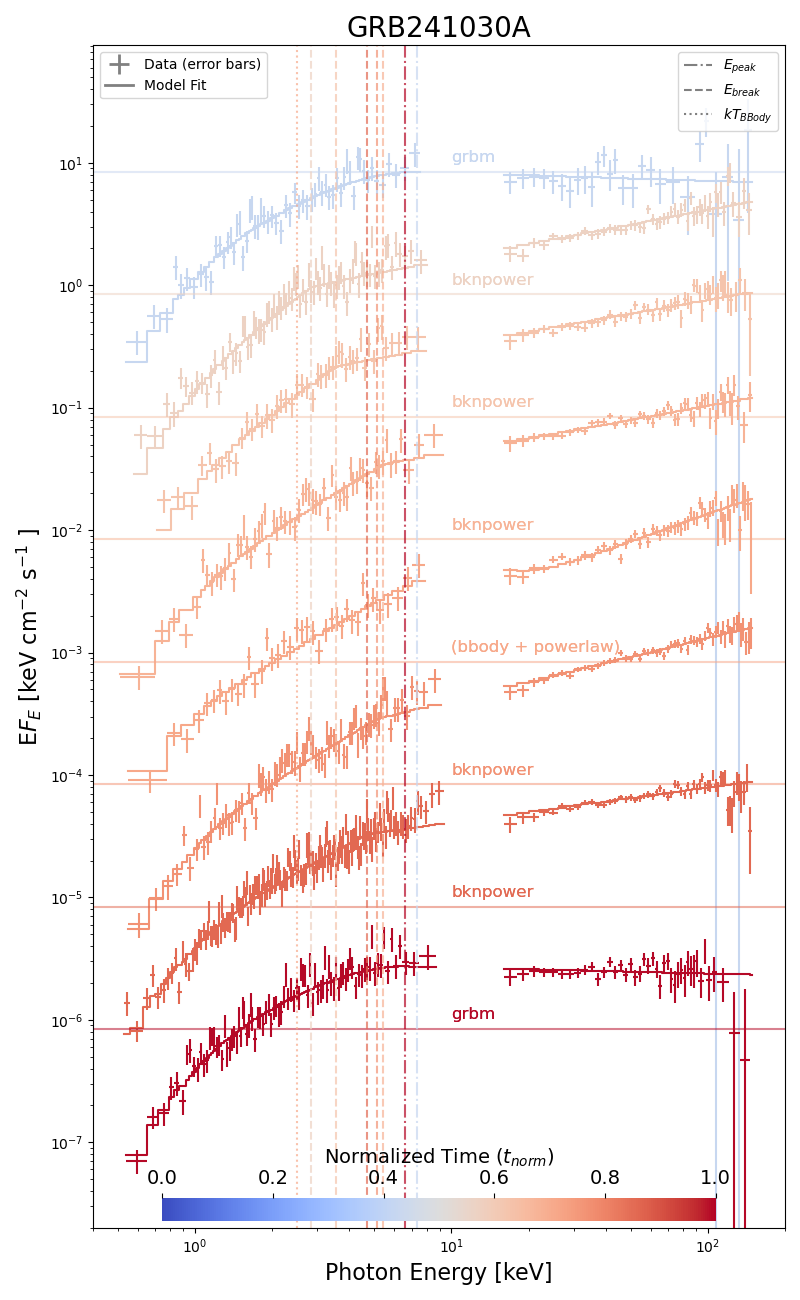}
    \caption{GRB 241030A: $\nu F_{\nu}$ spectra (SED) and best-fit model of 9 pulse intervals}
    \label{fig:241030spec}
\end{figure}

\section{Results}\label{results}
In the first part of this section, we discuss the parameter statistics and correlations among parameters of our pulses. In the second part, we discuss few GRBs with multiple pulses recorded well by both instruments.  
\subsection{Entire Sample}

\begin{deluxetable*}{ccc} \label{param-stats}
\label{tab:parameters}
\tablewidth{0pt}
\tablecaption{Gaussian fitted mean values and 1-$\sigma$ widths of the best-fit model parameters. } 
\tablehead{
\colhead{Parameter} & \colhead{Mean value} & \colhead{1$\sigma$ width}
}
\startdata
$\alpha_{\text{SPL}}$ &   -1.45 & 0.28\\
$\alpha_{CPL}$ &   -1.02 & 0.22\\
$\alpha_{1}$  &   -0.88 & 0.23\\
$\alpha_{2}$ &   -1.46 & 0.45\\
$\beta$ &   -2.37 & 0.35\\
$E_{\text{break}}$ &   6.03 keV & 3.43\\
\enddata
\end{deluxetable*}

\begin{figure} \label{models}
    \centering
    \includegraphics[width=0.47\textwidth]{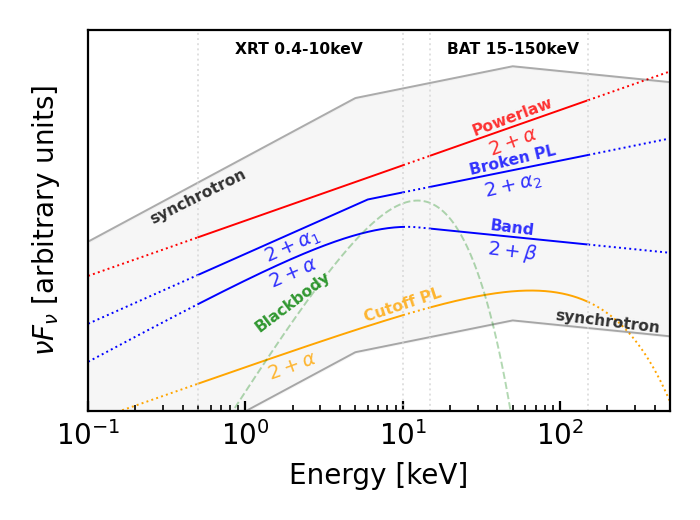}
    \caption{SED shapes of the power-law models used in this article. For BPL we denote index $\alpha_1$ and $\alpha_2$ below and above the break. For BPL with an SED peak observed, we denote indices $\alpha$ and $\beta$, with reference to the 'Band' shape. Quoted text refer to the percentage of best-fit models in our sample. There is also a possible subdominant blackbody. Grey shade represent an ideal fast-cooling synchrotron shape with arbitrary normalization.}
    \label{fig:model}
\end{figure}

All pulses can be well fit by one of the six PL based models displayed in Figure \ref{models} with $\chi^{2}_{red}$ $<$ 1.5. 
BPL model fit is required for 63\% (69/110) of the pulses. 
Higher energy spectra above the break energy are always softer than below the break, with $\alpha_{1} > \alpha_{2}$. 
The Gaussian-fit mean of $\langle \alpha_1 \rangle = -0.88$ and $\langle \alpha_2 \rangle = -1.45$ are very close to typical synchrotron indices, as shown in Table \ref{param-stats}. 
Spectra can be further categorized based on the presence of an SED peak, i.e., whether $\alpha_2$ is less than -2. 
Those without a peak are the low-energy break we were looking for, which appeared in 55/110(50\%) pulses. 
While other 14 pulses with an SED peak resemble the shape of classical `Band' spectra. 
We fit them using the Band function and obtained satisfactory results. 
%
%
Since $\alpha_{\text{Band}}$ reflects the spectrum below the peak, we include it in $\alpha_2$ statistics. 
%

\begin{figure}
    \centering
    \includegraphics[width=\linewidth]{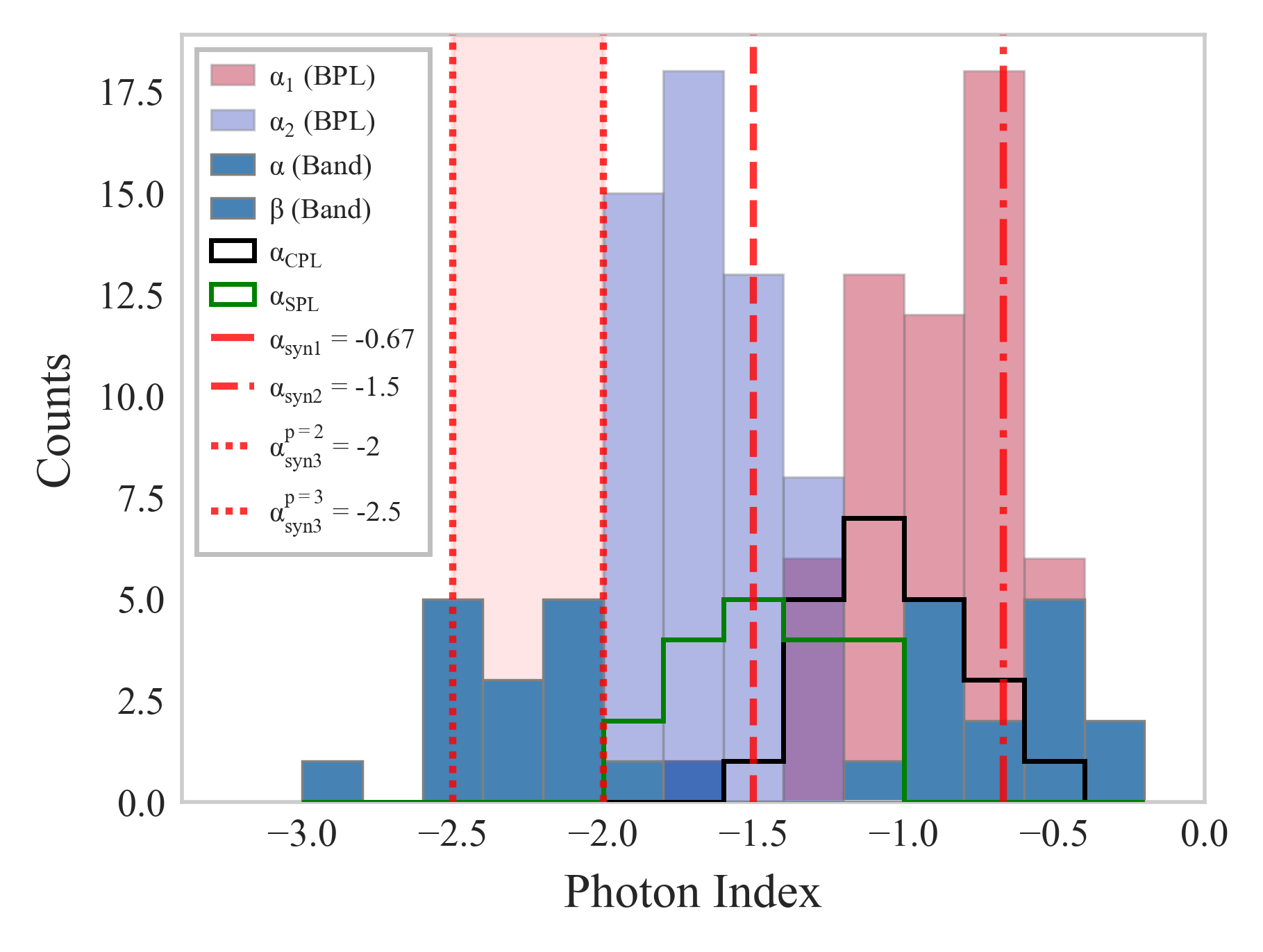}
    \caption{Distribution of power-law indices. Different color and linestyle refer to different models. Dashed, dotted lines and shades show typical synchrotron values and range. }
    \label{fig:a1a2b}
\end{figure}

CPL is the best fit for 19\% (21/110) of the pulses, while SPL is the best fit for 18\% (20/113).
We find $\langle\alpha_{\text{CPL}}\rangle = -1.04$ and $\langle\alpha_{\text{SPL}}\rangle = -1.45$. 
These values are consistent with the typical long GRBs low-energy index from \textit{\textit{Fermi}}-GBM \citep{2021ApJ...913...60P}: ‘comptonized’ $\alpha_{\text{COMP}} \simeq -1$ at below $E_{\text{peak}}$ and simple $\alpha_{\text{PLAW}}\simeq -1.6$. 
Distribution of above indices are shown on Figure \ref{fig:a1a2b}

The length of emission period in GRBs are not uniform \citep{2016Ap&SS.361..155H-BATT90}. 
We define a factor to better describe normalized mid time of the pulse interval within the prompt phase: \[
t_{\text{norm}} = \frac{t_{50,\text{interval}} - t_{5,\text{BAT}}}{T_{90,\text{BAT}}}
\]
where $t_{50,\text{interval}}$ represents the time of 50\% accumulated observed flux within the pulse interval. 
The addition of BAT 5\% flux time $t_{\text{5,BAT}}$ is intended to mitigate in some GRBs, the effect of long quiescent period between trigger and main emission phase. 
\begin{figure}
    \centering
    \includegraphics[width=0.48\textwidth]{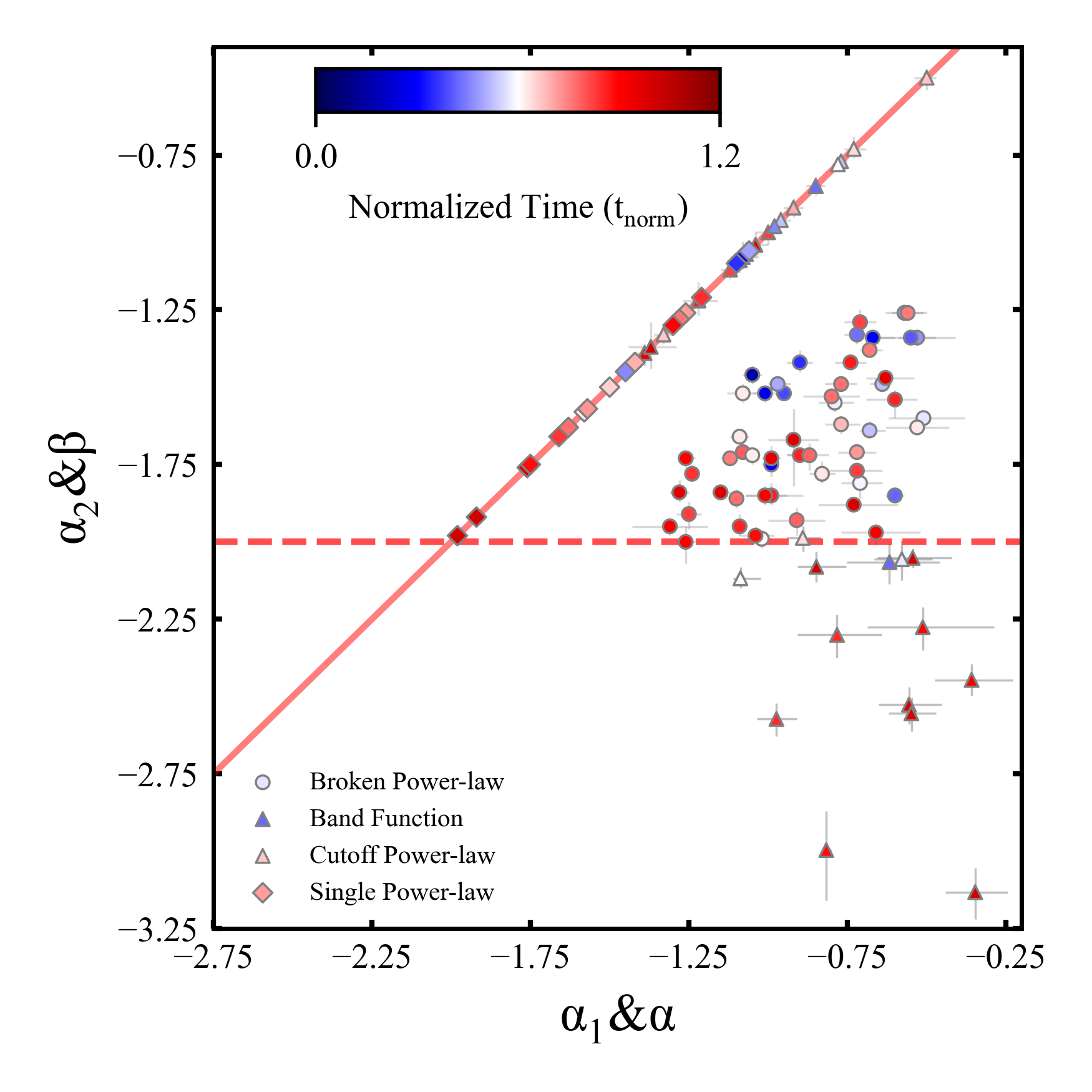}
    \caption{SPL pulses are plotted by diamonds while BPL are plotted by circles. CPL and Band pulses are plotted by triangles. $t_{\text{norm}}$ is shown by color-bar with earlier to blue and later to red. Error-bars show 1-$\sigma$ error ranges. Solid line represents the equality line while dashed line show $\beta=-2.0$. }
    \label{fig:gamma-gamma}
\end{figure}

Relations of indices along with $t_{\text{norm}}$ and 1$\sigma$ errors are shown in Figure \ref{fig:gamma-gamma}. 
An equality line reflects the same power-law index in both energy bands, with the CPL and SPL modeled pulses lying on this line. 
Models with a break are located in the lower-right region of the line, indicating a softer high-energy segment.
Whether best-fit by SPL/CPL or BPL, earlier pulses tend to be harder in both bands, being displayed in the top right of the figure. 
The ‘Band’ pulses show a harder $\beta$ in the early phase yet having a smaller sample size compared to other models.
This is in line with our expectations for hard-to-soft spectral behavior in GRBs.

\begin{figure} 
    \centering
    \includegraphics[width=0.48\textwidth]{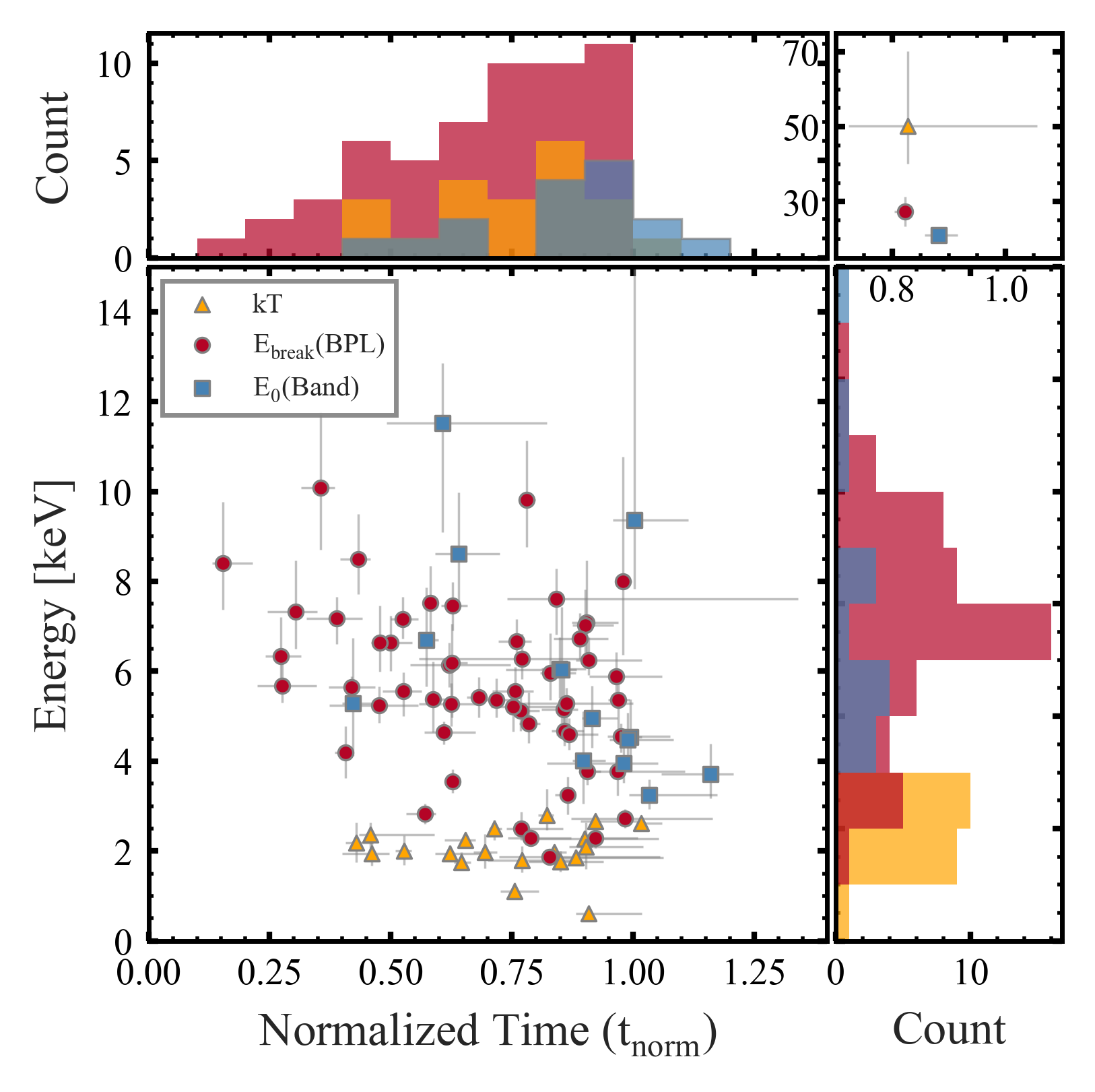}
    \caption{Relation between characteristic energy and normalized time. Right histogram shows energy distribution of break energy, Band $E_0$ and blackbody $\text{kT}$. Top histogram show frequency of characteristic energy found by normalized time. }
    \label{time-energy}
\end{figure}
Characteristic energies are shown in Figure \ref{time-energy}
BPL $E_{\text{break}}$ are mostly located near $\langle E_{\text{peak}} \rangle = 6.03\ \text{keV}$ and $\sigma = 3.43\ \text{keV}$. 
This supports the idea that prompt $E_{\text{break}}$ are intrinsically at energies below 20keV, which limited identification by instruments with higher sensitive bands \citep{Oganesyan2017ApJ...846..137O}. 
We calculate $E_{\text{peak}}$ by \(E_{\text{peak}} = (2+\alpha)E_{\text{cutoff}}\) for CPL and \(E_{\text{peak}} = (2+\alpha)E_{\text{0}}\) for Band. 
Band $E_{\text{peak}}$ has $\langle E_{\text{peak}} \rangle = 8.32\ \text{keV}$ and $\sigma = 4.85\ \text{keV}$. 
CPL $E_{\text{peak}}$ varies across different GRBs, yet fall within the range of $100$ to $500\ \text{keV}$. 
%

\begin{figure*}
\centering
\begin{minipage}{0.32\textwidth}
    \centering
    \includegraphics[width=\textwidth]{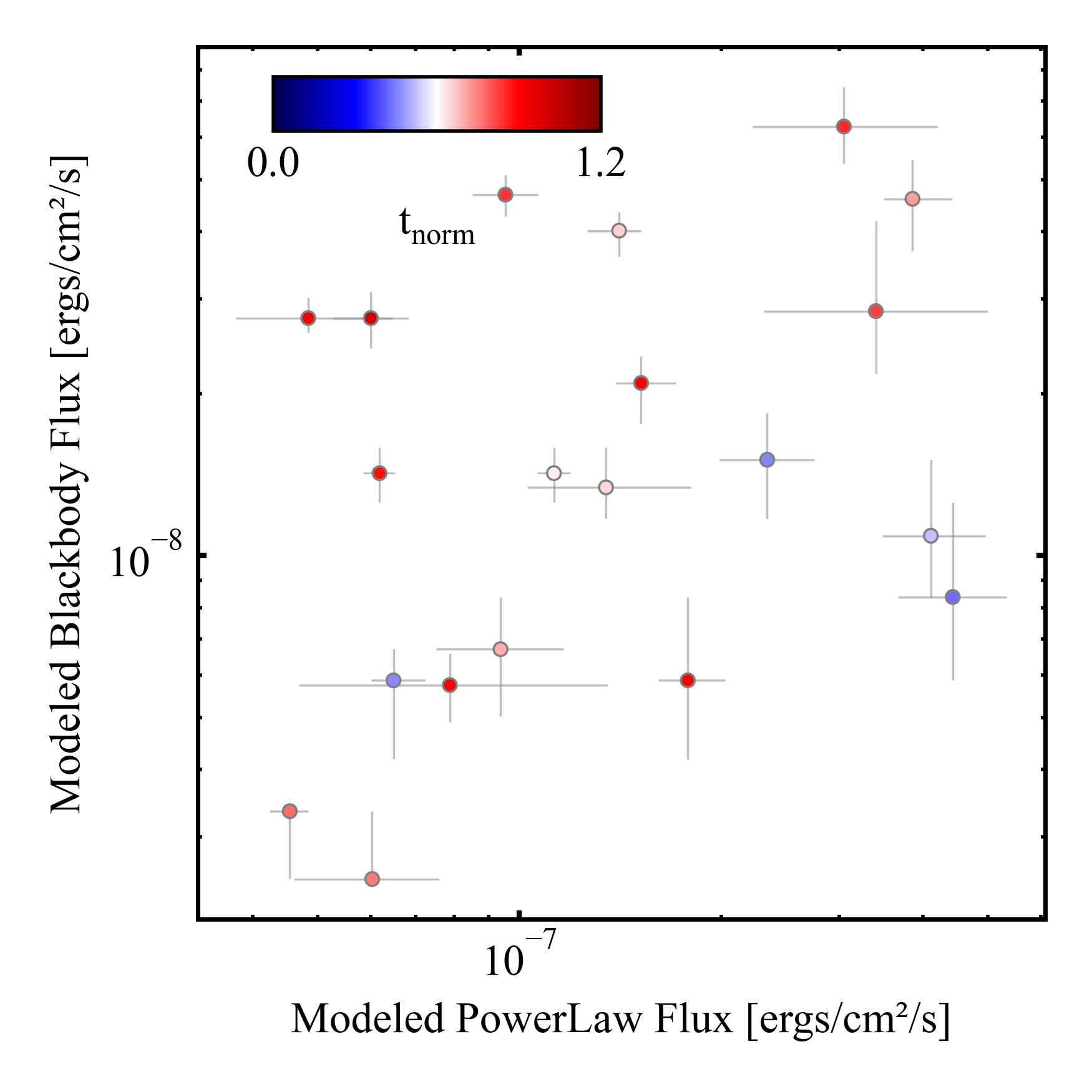}
\end{minipage}%
\begin{minipage}{0.32\textwidth}
    \centering
    \includegraphics[width=\textwidth]{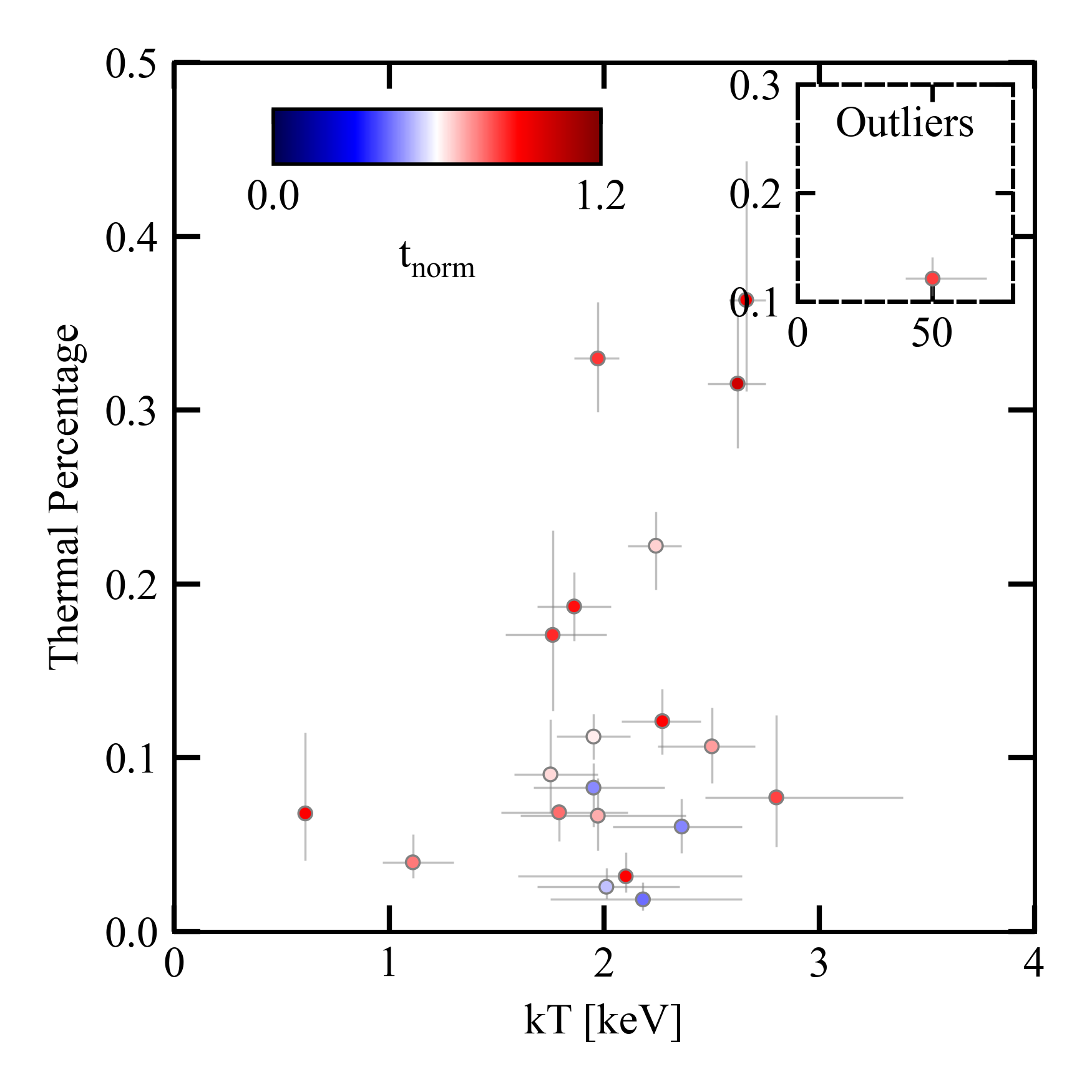}
\end{minipage}%
\begin{minipage}{0.32\textwidth}
    \centering
    \includegraphics[width=\textwidth]{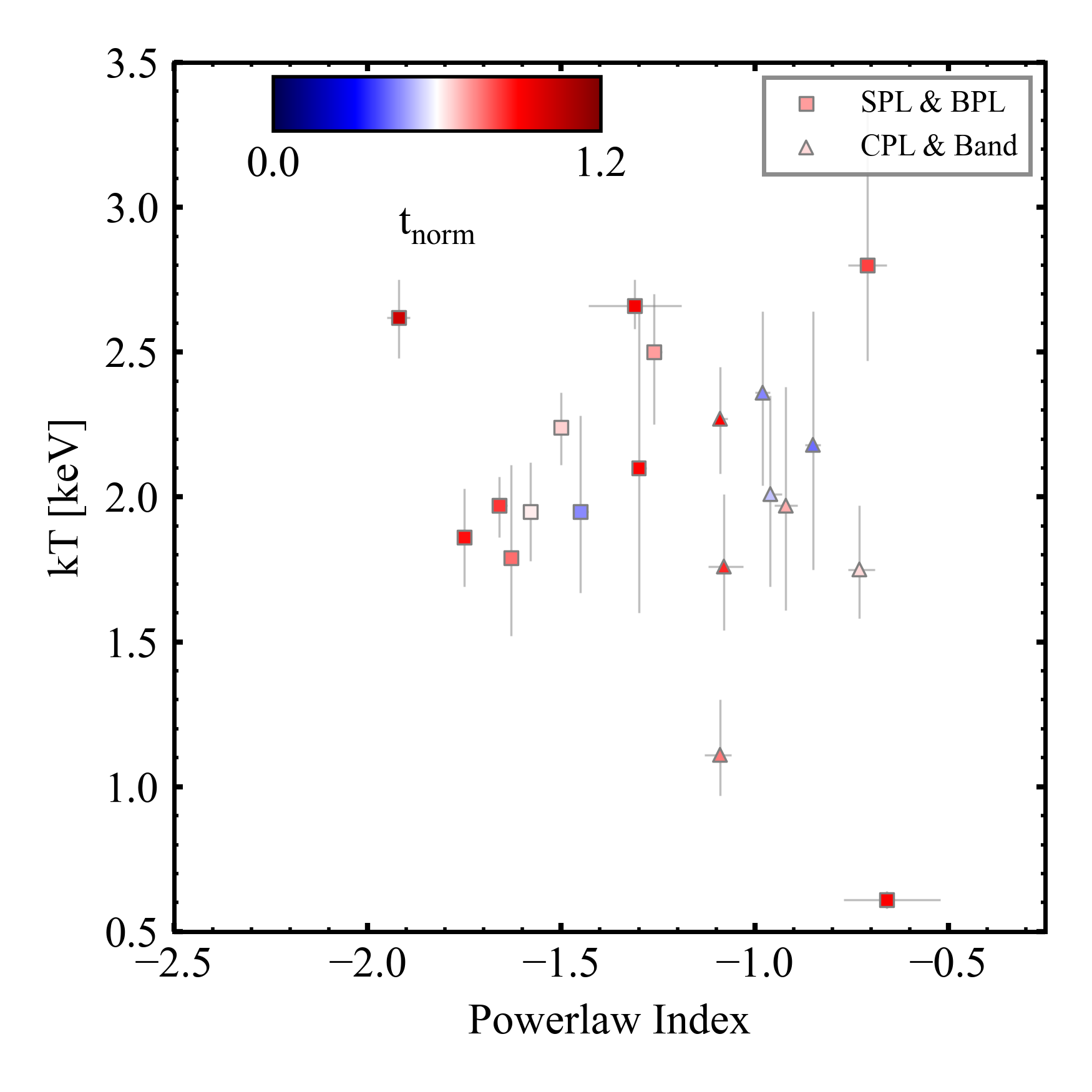}
\end{minipage}
\caption{Left panel: Modeled blackbody flux plotted against PL flux in 0.5-150keV. Central panel: Blackbody kT against modeled blackbody flux percentage in total flux. Right panel: Blackbody kT against photon indices in different models. No relations are found from blackbody parameters.}
\label{fig:BB_PL_kT}
\end{figure*}

An additive subdominant blackbody component is statistically favored 23 pulses. 
The observed kT values are predominantly below 3 keV as in Figure \ref{fig:BB_PL_kT}, and the thermal flux contributes less than 37\% (with a mean of 13\%) to the total emission in the 0.5-150 keV band. 
We conclude that such thermal components are subdominant and can be accommodated within our PL model interpretations.
With a blackbody component added to the PL models, most pulses are still best fitted with a break, compared to models such as BB+SPL or BB+CPL. 
Our conservative comparison criteria are sufficient to demonstrate that, even when considering the inclusion of a blackbody, the low-energy break remains a prominent feature in many pulses. 
%

We then present the distribution of parameters describing our BPL pulses in detail. 
By introducing the Band fit, we have categorized indices into $(\alpha_1,\alpha_2)$ for BPL and $(\alpha,\beta)$ for Band. 
%
%
Gaussian fit gives $\langle \alpha_{1} \rangle = -0.88$ with $\sigma = 0.23$, while the distribution shows a peak between $-0.6$ and $-0.8$, close to $\alpha_1^{\rm syn} = -0.67$.
The distribution gradually decrease towards $\sim -1.5$ on the lower end, while on the upper end, truncated near $\sim -0.5$.
This fits well with the synchrotron assumption that the low-energy segment with $\alpha_{1}$ corresponds to the fast-cooling synchrotron \citep{2018pgrb.book.....Z} $F_{\nu} \propto \nu^{\frac{1}{3}}$ spectral slope below $\nu_c$. 
Only a very small portion of intervals with $\alpha_{1} > -0.67$ exceeds the synchrotron line-of-death constraint, which is well within fitting error range. 
Our $\langle \alpha_{1} \rangle$ is steeper than the result $\langle \alpha_{1} \rangle = -0.66$ from the previous work \citep{Oganesyan2017ApJ...846..137O}. 
This results from taking the blackbody model into consideration, which provides an additional source of low-energy photons. 

Combining $\alpha_2$ with $\alpha$ in the statistics gives $\langle \alpha_{2} \rangle = -1.54$ with $\sigma = 0.31$.
The mean is close to the photon index produced by a fast-cooling synchrotron segment above $\nu_c$ and below $\nu_m$, which have $F_{\nu} \propto \nu^{-\frac{1}{2}}$. 
The softer end of $\alpha_2$ distribution contains more pulse samples, truncates at the defined upper limit of $\alpha_{2} = -2.0$. The harder end extends towards the common observed index in Gamma-ray band: $\alpha \simeq -1$. 
The Band $\alpha$ is noticeably harder, showing a distribution centered at $\alpha_{\text{Band}}=-0.75$. 
This is because the smoothly broken shape of the Band function at low energies requires a harder low-energy spectral index.
%

Band function $\beta$ is concentrated at $\langle \beta \rangle = -2.37$ with $\sigma = 0.32$.
The fitted range between $-2.5 <\beta < -2.0$ align well with fast-cooling synchrotron slope with $F_{\nu} \propto \nu^{-p/2}$ at above $\nu_m$, taking $p = 2 \sim 3$. 
%

Results of $\alpha_1,\alpha_2$ and $\alpha$ are consistent with the simple/cutoff PL index distribution from previous \textit{Swift} BAT catalogs \citep{Sakamoto2011ApJS..195....2S,Lien2016ApJ...829....7L} and the PLAW/COMP low-energy index distribution from \textit{Fermi}-GBM catalog. 
$\beta$ falls well within the range of GBM BAND/SBPL model high-energy indices although with a limited sample of 17 pulses. 
Relation between the break energies and spectral indices are displayed in Appendix \ref{distribution-of-parameters}. 
%

Figure \ref{fig:index_diff} shows the difference between the two spectral indices $\Delta S = \alpha_2-\alpha_1$ for BPL models. 
We obtain a Gaussian mean $\langle \Delta S \rangle = -0.78 \, (\sigma = 0.21)$, reflecting that across all pulses the two indices soften in a similar manner, and the degree of spectral change between the two power-law segments does not vary significantly. 
Notably, not only do $\alpha_1$ and $\alpha_2$ values agree with the synchrotron prediction, but mean index difference $\langle \Delta S\rangle$ is also consistent with the expected synchrotron fast-cooling break at $\nu_c$ which has $\alpha^{\rm syn}_2 - \alpha^{\rm syn}_1 = -0.83$. 

From the widely observed BPL $E_{\text{break}}$ together with the distributions of $\alpha_1$, $\alpha_2$, and $\Delta S$, we interpret this low-energy break as the synchrotron cooling frequency $\nu_c$, while the $E_{\text{peak}}$ in CPL and Band models is attributed to the minimum synchrotron frequency $\nu_m$. 
$\nu_c$ remains within roughly one order of magnitude. With an index distribution near $\alpha_2 \simeq -1.5$, these spectra should be the case of fast-cooling regime of $\nu_c < \nu_m$. Whereas $\nu_m$, derived from CPL or Band fits, exhibits a wider range of values from few keV to hundreds of keVs. For SPL spectra, $\nu_c$ lies on the lower-energy side while $\nu_m$ is on the higher-energy side of the observable band. Indices distributes near $\alpha \simeq -1.5$, which is also consistent with the fast-cooling model. 

%
%
\begin{figure}
    \includegraphics[width=0.47\textwidth]{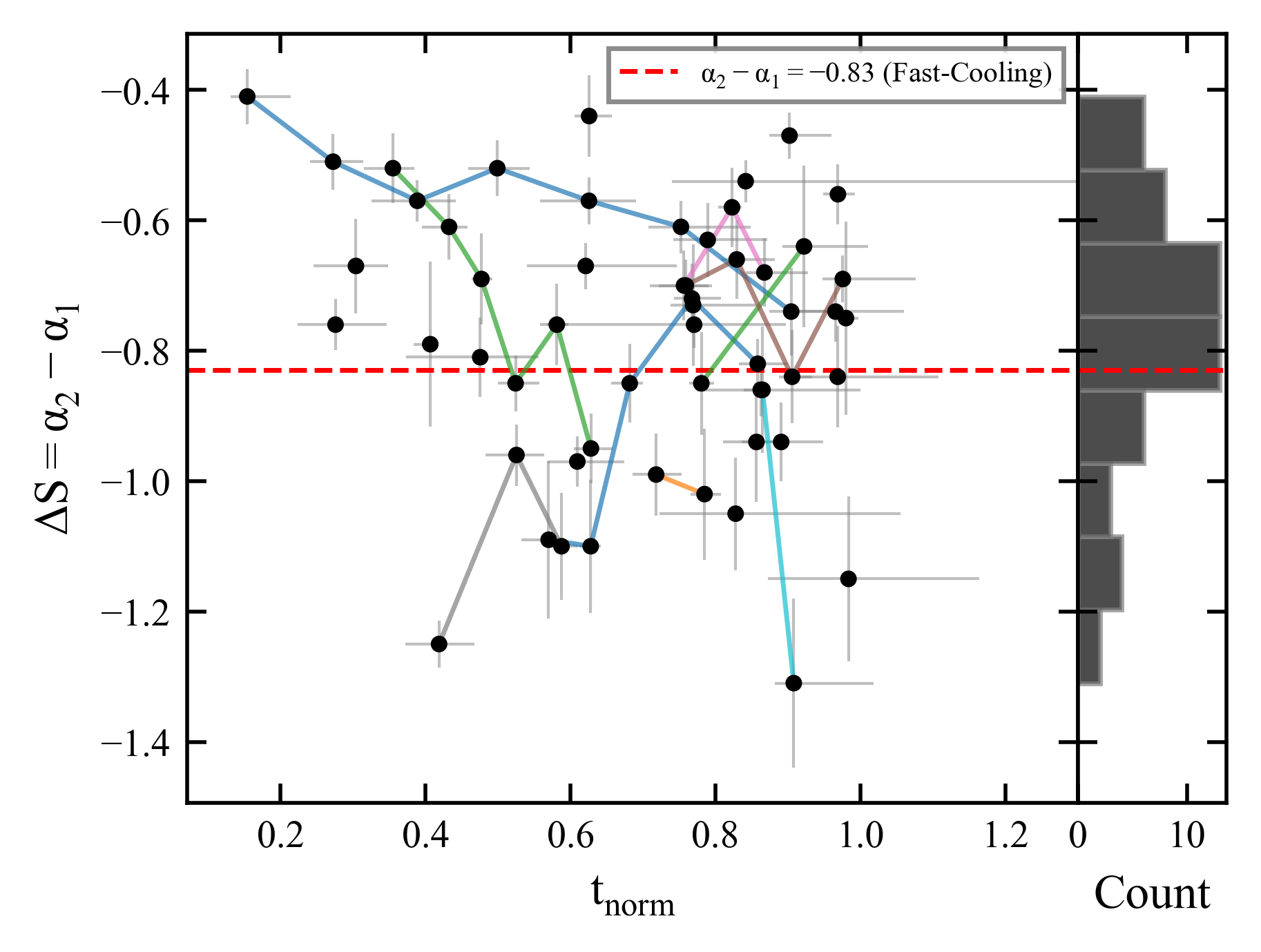}
    \caption{Distribution of spectral index differences. Typical fast-cooling synchrotron index differences are shown with dotted lines. Different colored lines represent pulses in one same GRB. }
    \label{fig:index_diff}
\end{figure}

%

\subsection{Individual GRBs}
There are few GRBs that contain multiple pulses with ambient photon counts: GRB 241030A, GRB 190604B, GRB 140619A, GRB 140512A, GRB 121123A, GRB 110205A, GRB 110102A, GRB 100728A and GRB 070616. In these bursts, we could examine how parameters vary by normalized time. 

$E_{\text{break}}$ have insignificant variation throughout time-resolved pulses. 
In particular, GRB 070616, for which early-time X-ray spectra are available, shows $E_{\text{break}} = 8.4^{+1.4}_{-1.0}\ \text{keV}$ at $t_{\text{norm,1}} = 0.15$ (subscript denotes the pulse order), slightly decreasing over time to $E_{\text{break}} = 5.2^{+0.4}_{-0.6}\ \text{keV}$ at $t_{\text{norm,6}} = 0.75$, and then increasing to $E_{\text{break}} = 7.1^{+1.4}_{-0.7}\ \text{keV}$ in the last pulse at $t_{\text{norm,7}} = 0.90$. 
In GRB 241030A $E_{\text{break}}$ first increases from $2.9^{+0.2}_{-0.2} \text{keV}$ to $5.4^{+0.5}_{-0.5} \text{keV}$ and then decreases to $4.7^{+0.4}_{-0.3} \text{keV}$. 
Although pulse 5 is classified as BB+SPL under our selection criteria, a BB+BPL model provides a modest improvement by $\Delta BIC \approx 2.5$. In BB+BPL case there are $E_{\text{break}} = 2.4^{+0.6}_{-0.5}\ \text{keV}$, $\alpha_1=-0.72$ and $\alpha_2=-1.39$, which are consistent with the expected indices trends. 
GRB 140619A, GRB 110102A and GRB 110205A show similar behavior, with no significant $E_{\rm break}$ variation during the observed period. 
GRB 140512A stands out as the only case exhibiting a relatively larger variation in $E_{\text{break}}$, increasing from $5.6^{+0.6}_{-0.4}\ \text{keV}$ to $27.3^{+4.0}_{-3.9}\ \text{keV}$ at $t_{\text{norm,2}} = 0.80$, before decreasing to $4.6^{+0.4}_{-0.3}\ \text{keV}$ at $t_{\text{norm,3}} = 0.84$. 

$E_{\text{peak}}$ shows clear decaying trends in contrast. 
Specifically, in GRB 121123A, $E_{\text{peak}}$ drops from $127.4^{+14.4}_{-12.0}\ \text{keV}$ to $45.5^{+5.1}_{-4.6}\ \text{keV}$ across all five pulses; in GRB 100728A, the CPL pulses at $t_{\text{norm}} = 0.79–0.99$ show $E_{\text{peak}}$ decreasing from $206.7^{+41.6}_{-30.4}\ \text{keV}$ to $43.0^{+9.4}_{-7.3}\ \text{keV}$; and in GRB 100725B, $E_{\text{peak}}$ evolves from $66.8^{+8.1}_{-6.8}\ \text{keV}$ to $6.9^{+1.2}_{-1.0}\ \text{keV}$ between $t_{\text{norm}} = 0.81$ and $1.36$. 

In bursts having consecutive BPL pulses, both $\alpha_1$ and $\alpha_2$ show a general decrease from early to late times, though with some variations along the way. In GRB 241030A, GRB 190604B, GRB 100728A, and GRB 100725B, the best-fit models of pulses evolve from BPL/SPL at earlier times to Band/CPL at later times, with a significant decrease in the flux contribution from the high-energy band. 

In summary, spectral shape between pulses exhibits a hard-to-soft evolution trend. 
Between consecutive BPL pulses, $E_{\text{break}}$ varies within less than one order of magnitude. For bursts with consecutive CPL peaks observed, $E_{\text{peak}}$ shows a decreasing trend. According to the correspondence with the synchrotron model, our result reflects that $\nu_c$ remains nearly constant with normalized time, whereas $\nu_m$ exhibits a decreasing behavior. 
However, we have to keep in mind that these results come from a very limited sample, thus we need to be cautious before applying them as universal conclusions. 

\section{Conclusion and Discussion} \label{summary}
The existence of an intrinsic low-energy break around few keV had been identified by few previous studies. 
The introduction of the break resolved the issue of X-ray hardening commonly observed at low energies.
However these studies have either focused on a single event, or required well-resolved temporal features during the joint-observation interval. 
In this work, we relax the selection criteria, extending the sample to include GRBs with only one or a few pulses meeting the SNR5 requirement. 
In addition, we update the dataset to cover 20 years of \textit{Swift} BAT–XRT joint observations, extending through March 2025.

We confirm that a single PL model (with break or cutoff), considering a subdominant thermal component, can well describe the prompt spectrum from 150 keV down to 0.5 keV. 
We also find popular existence of the break in more than half of sample pulses and located $E_{\text{break}}$ down to $\sim6$keV, as well as common spectral index evolution. 
By allowing a possible blackbody component, we obtain an $\alpha_{1}$ range below $E_{\text{break}}$ that does not exceed the line-of-death. 
Evolution of $E_{\text{peak}}$ are seen in few sample GRBs. 
The distribution of $\alpha_{1},\alpha_2$ and $\beta$ are in very good agreement with fast-cooling 
synchrotron slopes below and above $\nu_c$, $\nu_m$, which is applicable to a broad range of GRB pulses. 

In the ideal case, only typical index values should be observed. 
However for both $\nu_c$ and $\nu_m$, the physical processes associated with the breaks occur over a finite energy width rather than being perfectly sharp. 
As a result, the difference between the fitted breaks and the true frequency can introduce additional photons on either side, slightly modifying the spectral indices.
Also, a marginally fast-cooling regime leading to spectrum curvature \citep{2011A&A...526A.110D,2013ApJ...769...69B,2014NatPh..10..351U} was also suggested as an explanation. 
Varying electron injection which can lead to such non-standard values and non-power-law (curved) cooling slopes, is also discussed in \citep{2019ApJ...886..106P}. 
Varying electron injection and magnetic field, which can lead to such non-standard values and non-power-law (curved) cooling slopes, are also discussed in \citep{2019ApJ...886..106P}. 
Our result $\langle \alpha_1 \rangle = -0.88$ from pulse spectra from weaker GRBs and late-epoch emission confirms the possibility of these theories. 
This study thus provide evidence that the GRB prompt emission is predominantly produced by synchrotron radiation from a single central origin, and that the different spectral features observed at various times are the result of spectral evolution. 
The different characteristic energies and slopes of the observed spectra are likely caused by the evolution of the synchrotron $\nu_c$ and $\nu_m$, combined with the limitations of the observing energy band. 
A lack of correlation between spectral parameters and \( t_{\text{norm}} \) suggest a chaotic and heterogeneous nature of the GRB central engine. 

However, there are limitations worth noting. Without involving high energy instruments in typical GRB bands, constraint on $E_{\text{peak}}$ is limited. We did not adopt more complex model to characterize $E_{\text{break}}$ and $E_{\text{peak}}$ at the same time. 
Intrinsic absorption is also difficult to constrain, introducing uncertainty in the low-energy part of spectra. 
In the future, it will be worthwhile to explore larger samples with broad band observations. 
State-of-the-art, large-FOV instruments covering the soft X-ray band—such as \textit{SVOM}-ECLAIRs \citep{2009AIPC.1133...25G,2014SPIE.9144E..24G} and \textit{EP}-WXT \citep{2022hxga.book...86Y}, together with other gamma-ray instruments, would hopefully  observe more GRB with simultaneous detections down to few keVs. Based on such data, constraining both $E_{\text{peak}}$ and $E_{\text{break}}$ with fitting functions that incorporate two characteristic energies (such as BCPL or Bandcut) will provide stronger insights into the nature of the prompt emission.


%


\begin{acknowledgments}
``This research has made use of data, software and/or web tools obtained from the High Energy Astrophysics Science Archive Research Center (HEASARC), a service of the Astrophysics Science Division at NASA/GSFC and of the Smithsonian Astrophysical Observatory's High Energy Astrophysics Division." This work is supported by the National Natural Science Foundation of China (Projects 12373040,12021003) and the Fundamental Research Funds for the Central Universities.
\end{acknowledgments}




%
\facilities{\textit{Swift}(XRT and BAT)}

\software{HEASoft \citep{2014ascl.soft08004N},
          astropy \citep{2013A&A...558A..33A,2018AJ....156..123A,2022ApJ...935..167A},  
          XSPEC \citep{2013RMxAA..49..137F}, 
          pyXspec \citep{1996A&AS..117..393B}
          }

\clearpage
\appendix
\section{distribution of parameters}\label{distribution-of-parameters}

\begin{figure*}[b]
\centering
\begin{minipage}{0.48\textwidth}
    \centering
    \includegraphics[width=\textwidth]{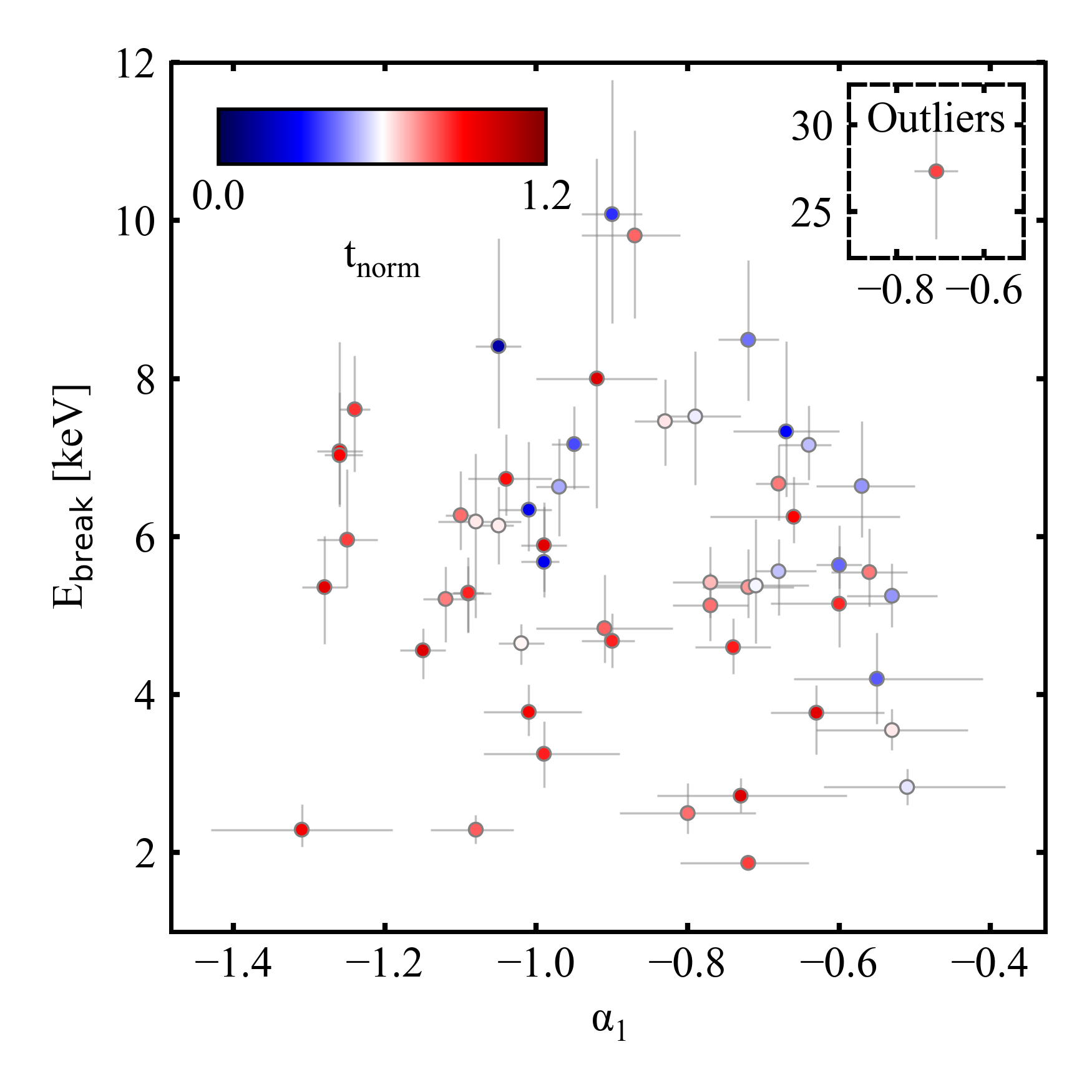}
    \label{fig:Eb-alpha1}
\end{minipage}
\hspace{0.01\textwidth}
\begin{minipage}{0.48\textwidth}
    \centering
    \includegraphics[width=\textwidth]{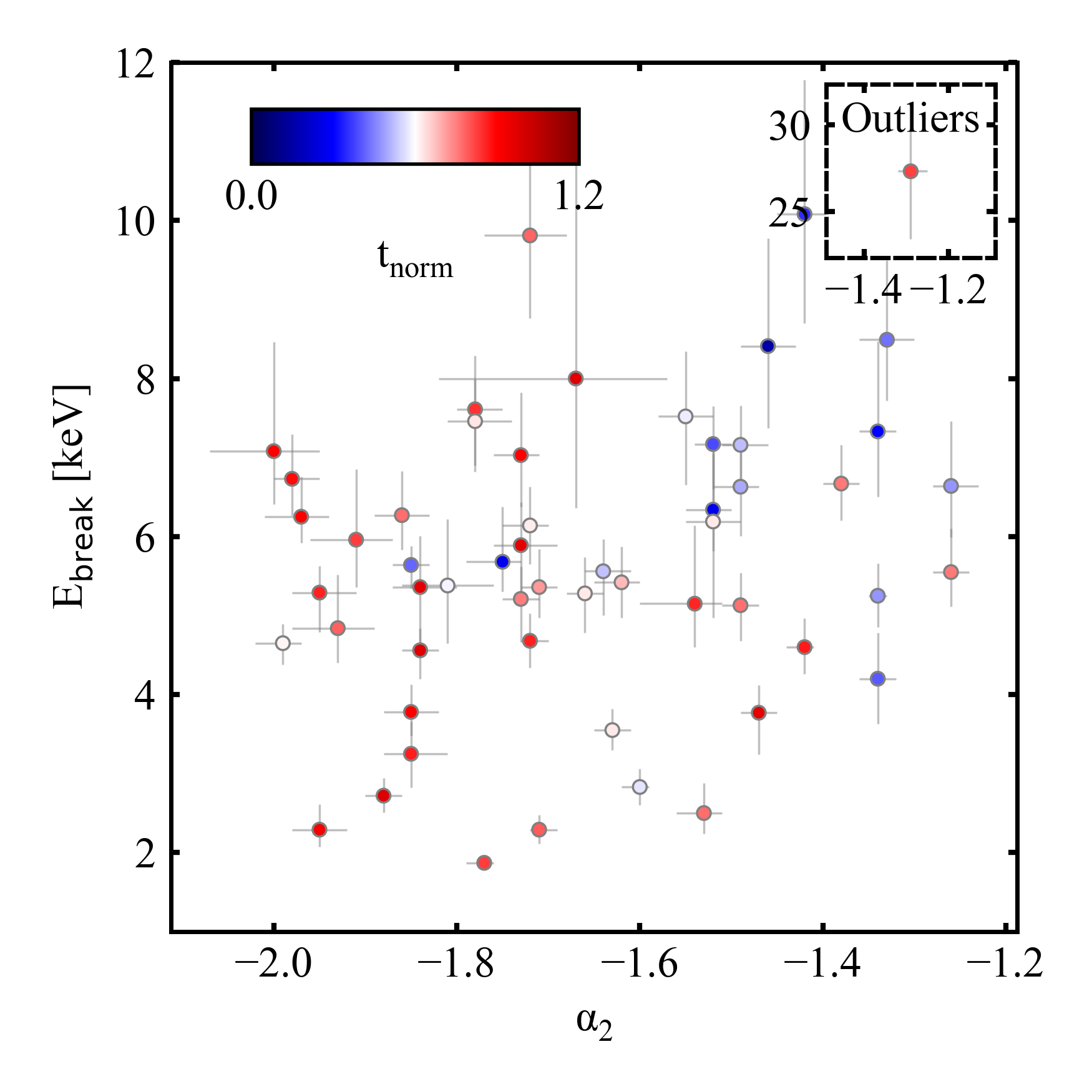}
    \label{fig:Eb-alpha2}
\end{minipage}

\vspace{0.02\textwidth}

\begin{minipage}{0.48\textwidth}
    \centering
    \includegraphics[width=\textwidth]{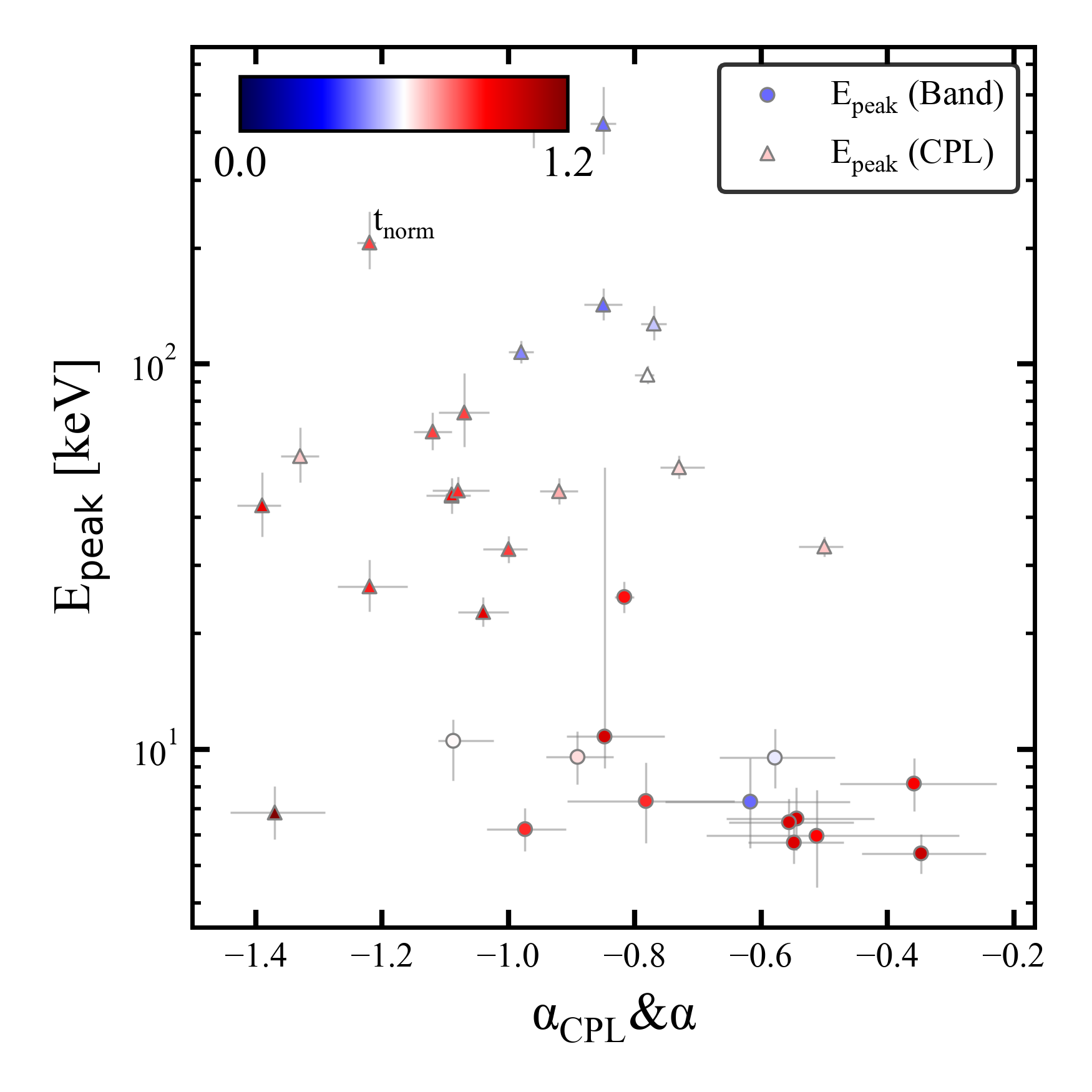}
    \label{fig:Ep-alpha}
\end{minipage}
\hspace{0.01\textwidth}
\begin{minipage}{0.48\textwidth}
    \centering
    \includegraphics[width=\textwidth]{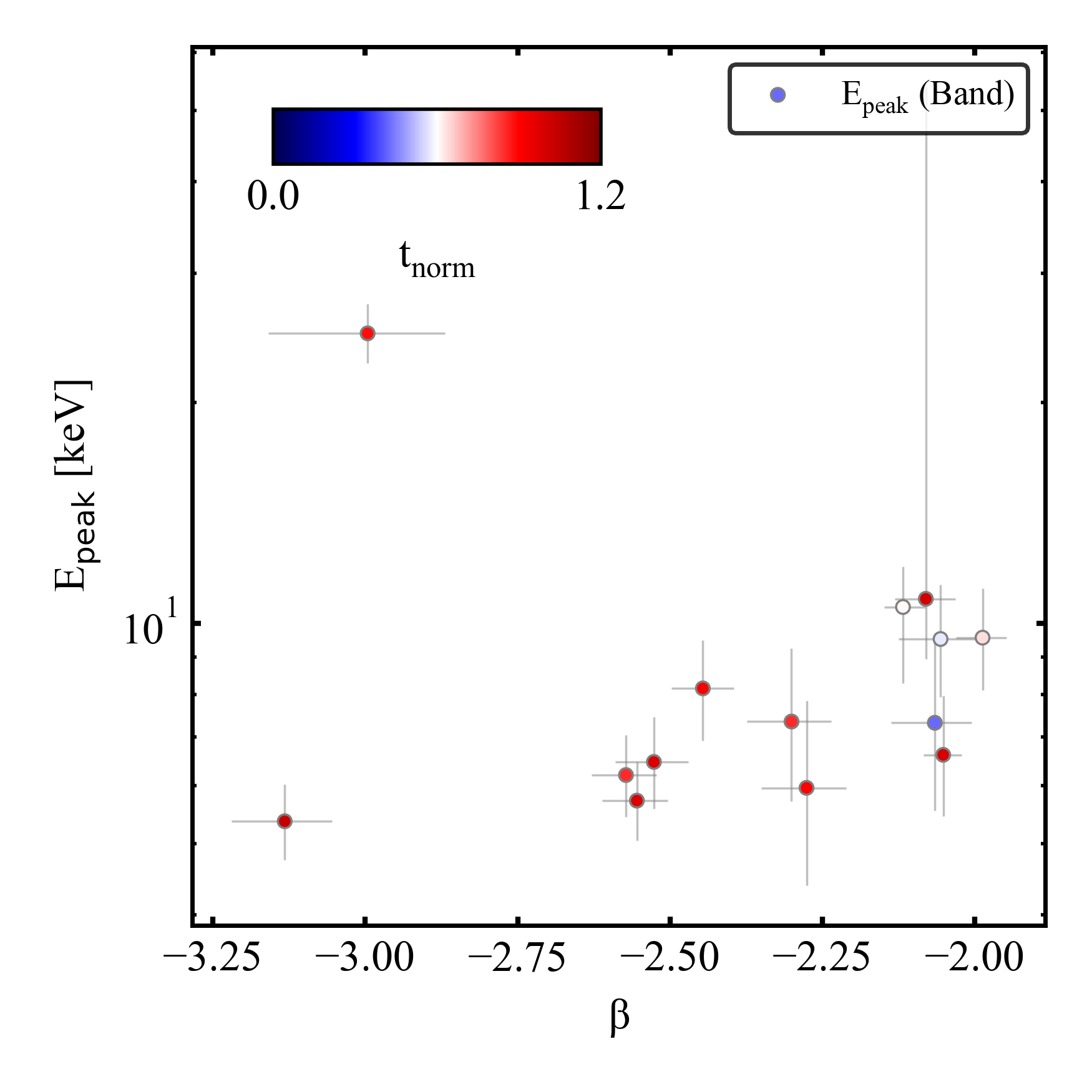}
    \label{fig:Ep-beta}
\end{minipage}

\caption{
Relations between spectral indices and characteristic energies. 
Color map indicates $t_{\text{norm}}$ of pulses. 
Error bars represent $1\sigma$ range of the fitted parameters. 
Top row: $E_{\text{break}}$ vs. $\alpha_1$ and $\alpha_2$. 
Bottom row: $E_{\text{peak}}$ vs. $\alpha$ and $\beta$.
}
\label{fig:Eb_Ep_index}
\end{figure*}
\clearpage


\section{resolved GRB light curves} \label{all-lc}
In each of the following figures, top panel is the corrected count rate light curve in both bands, with left Y-axis represent XRT counts, and X-axis represent BAT counts. Central panel contains the fitted spectral parameters for each pulse interval. Bottom panel contains the fitted characteristic energies within the pulse interval. The error bars on the Y-axis represent the $1\sigma$ range of the fitted parameters, while those on the X-axis correspond to the times of 5\% and 95\% cumulative flux within each interval.
\begin{figure}[h]
    \centering
    \includegraphics[width=1\linewidth]{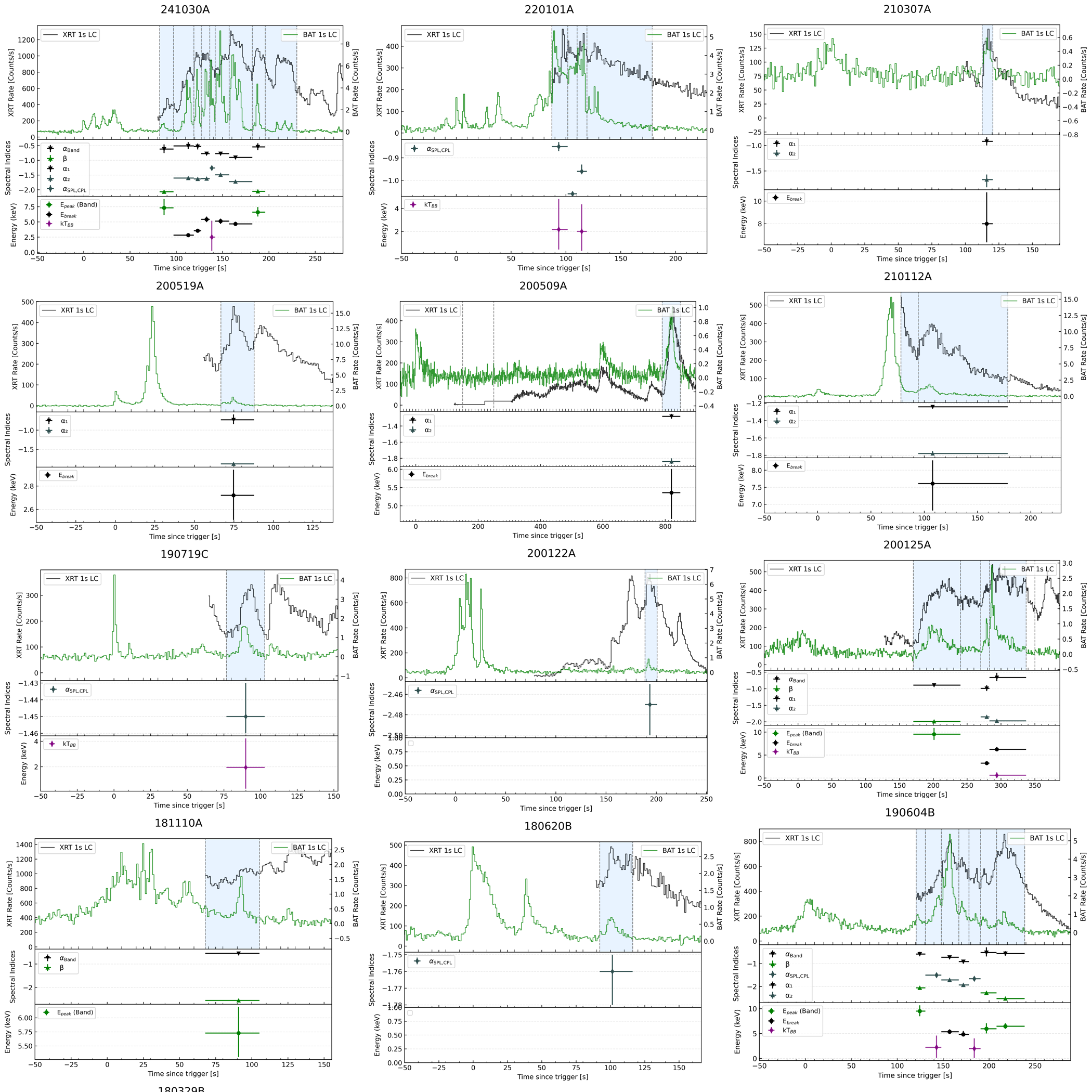}
    \label{fig:placeholder}
\end{figure}
\begin{figure}[h]
    \centering
    \includegraphics[width=1\linewidth]{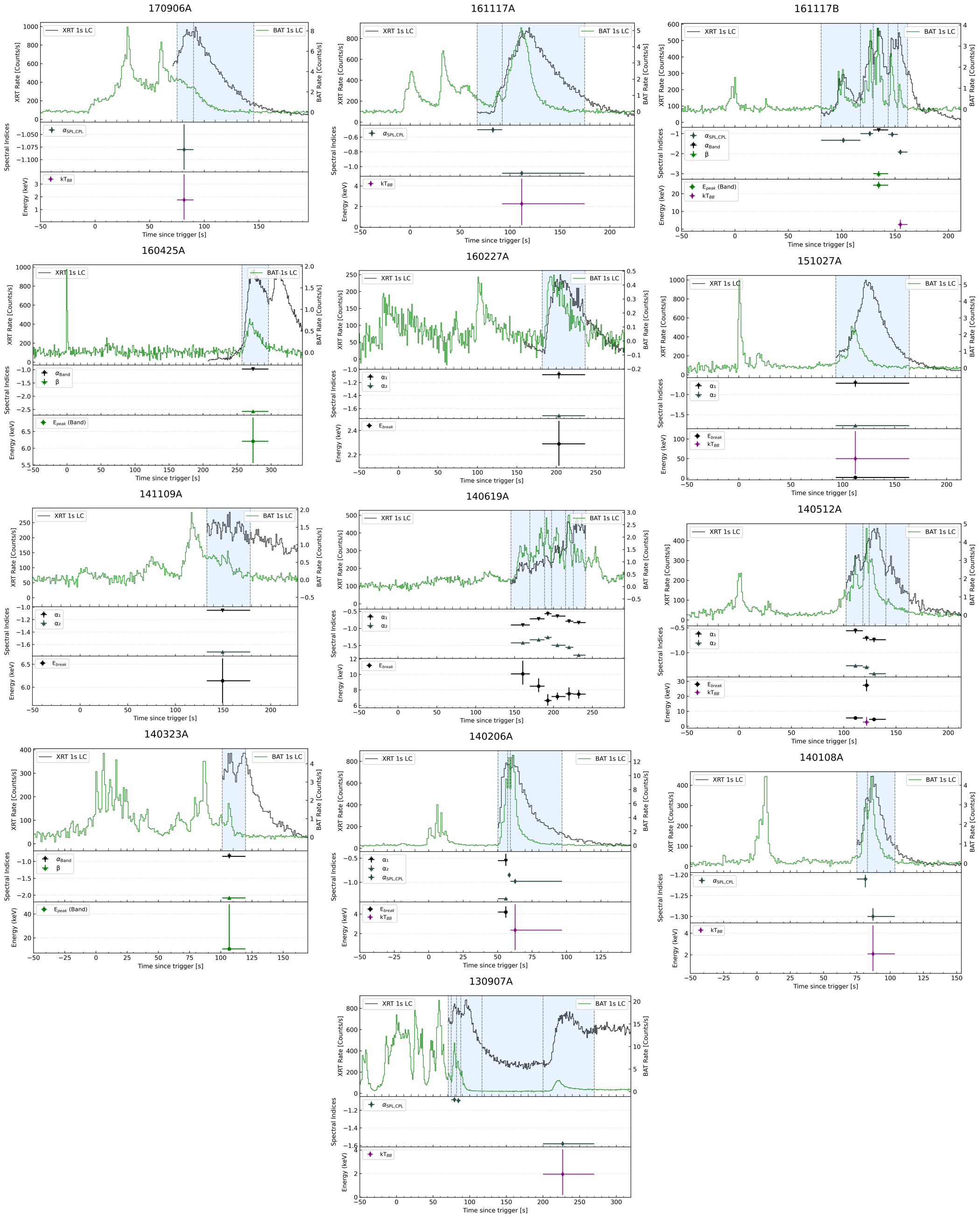}
    \label{fig:placeholder}
\end{figure}
\begin{figure}[h]
    \centering
    \includegraphics[width=1\linewidth]{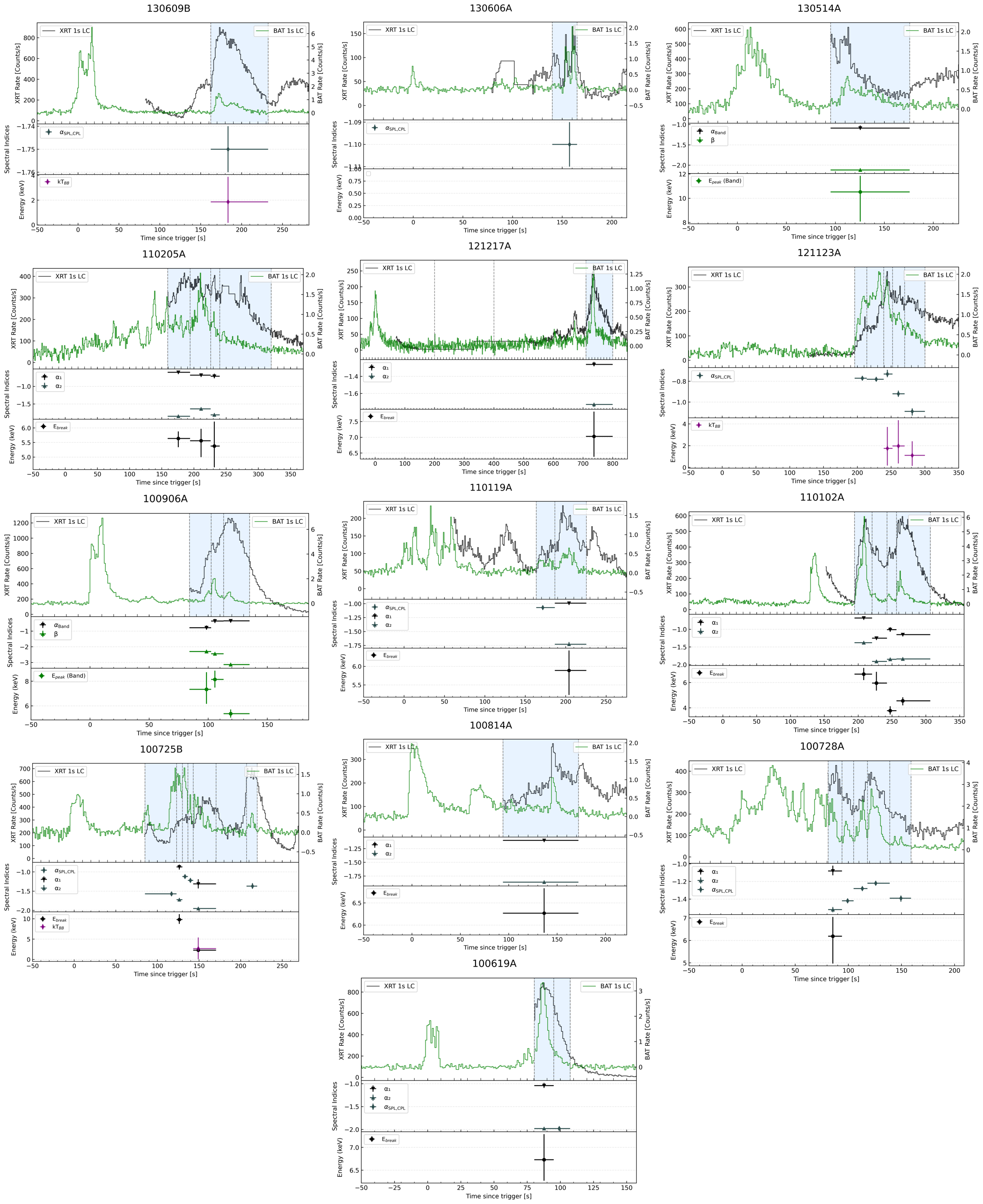}
    \label{fig:placeholder}
\end{figure}
\clearpage
\begin{figure}[h]
    \centering
    \includegraphics[width=1\linewidth]{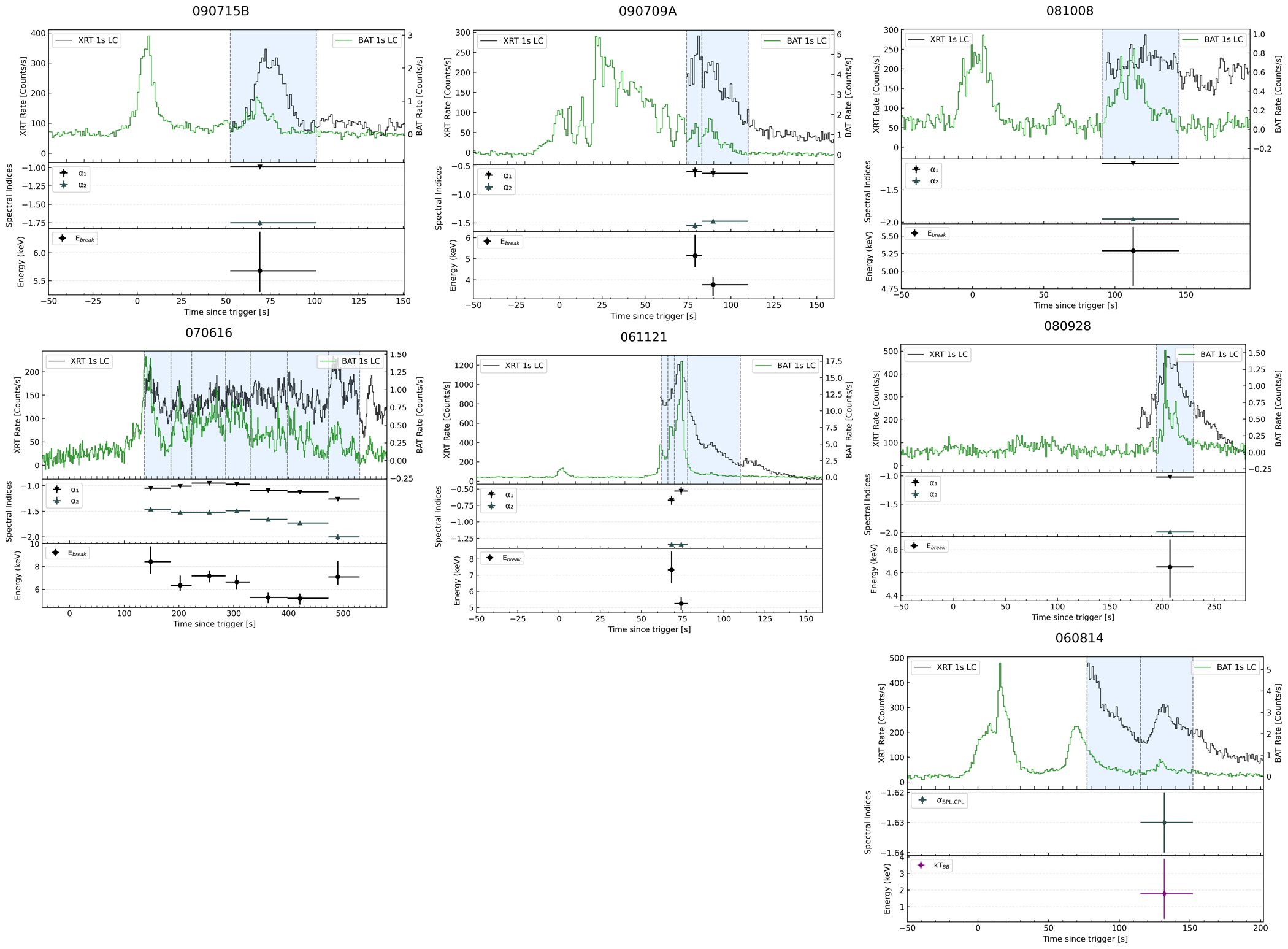}
    \label{fig:placeholder}
\end{figure}

\section{resolved GRB spectra} \label{all-spec}
Color represent the relative occur time for the pulse. Earlier pulses are more to the blue color, while later pulses are more to the red. 
Solid step lines shows the best-fit model for the pulse. 
A fixed offset of 0.1x in fluxes are applied to each pulse, with horizontal dashed reference lines of model predicted power at 10keV after applying the offset. 
Break Energies are illustrated by dashed lines. Cutoff Energies are illustrated by dotted-dashed lines. 
\begin{figure}[h]
    \centering
    \includegraphics[width=1\linewidth]{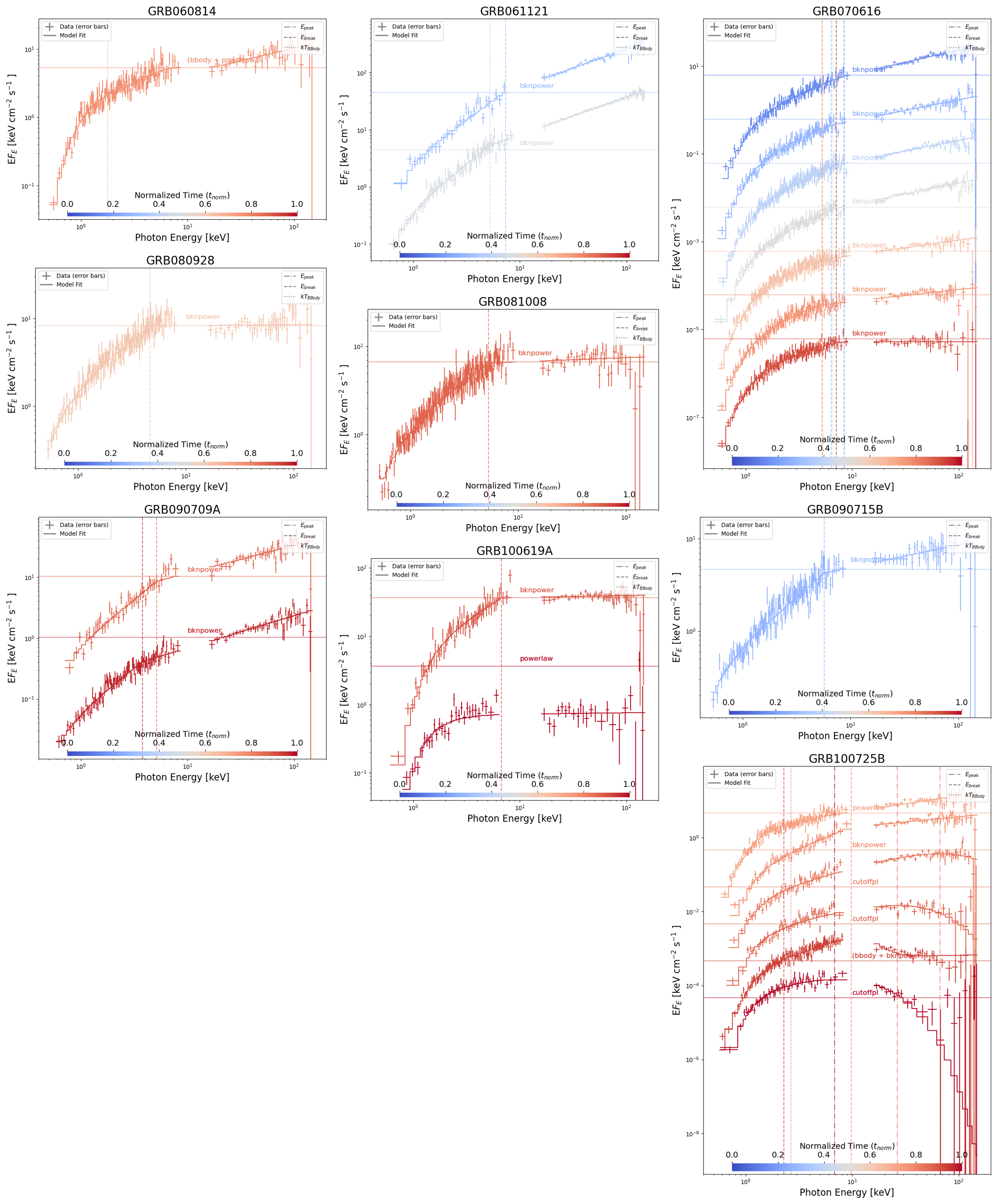}
    \label{fig:placeholder}
\end{figure}
\begin{figure}[h]
    \centering
    \includegraphics[width=1\linewidth]{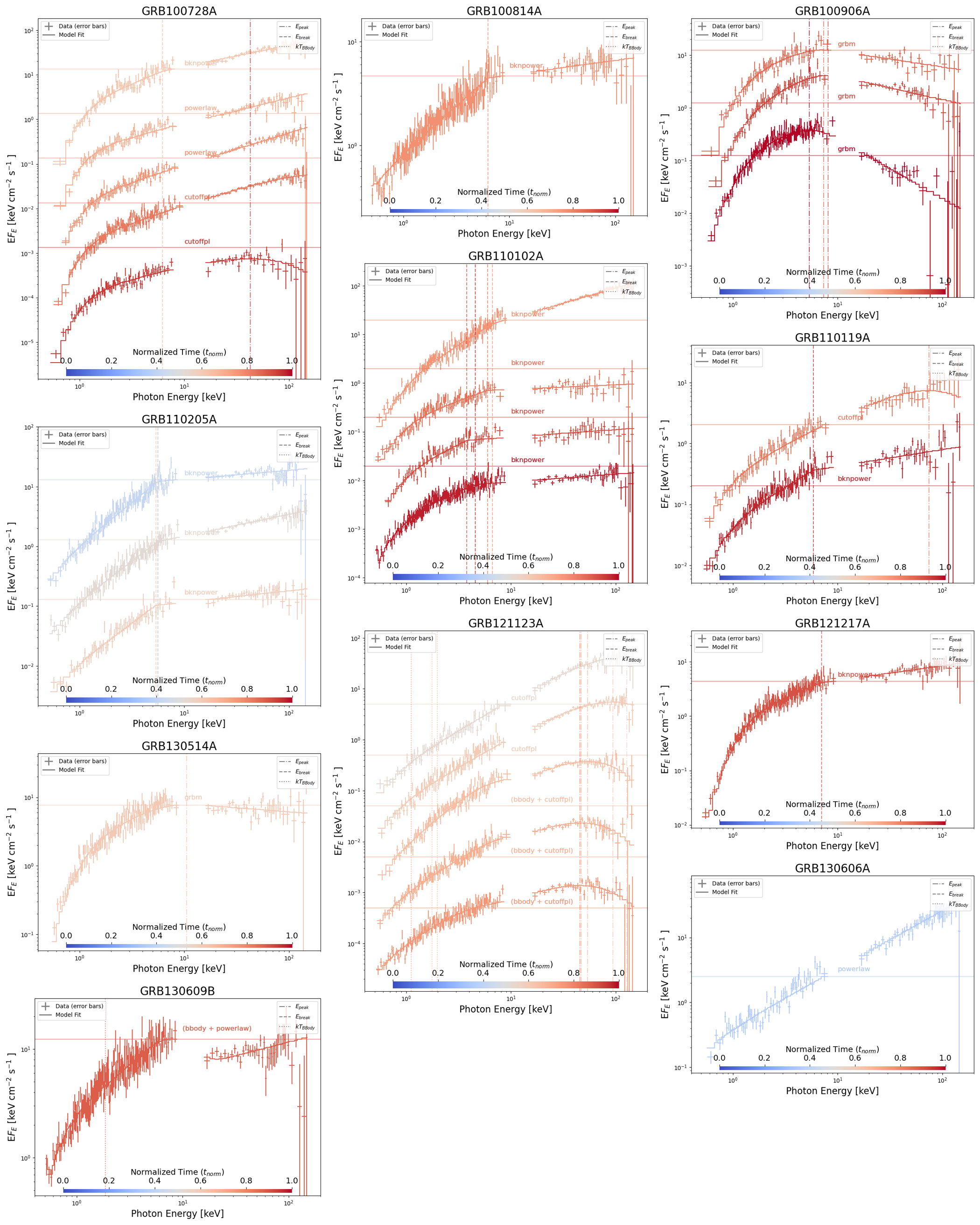}
    \label{fig:placeholder}
\end{figure}
\begin{figure}[h]
    \centering
    \includegraphics[width=1\linewidth]{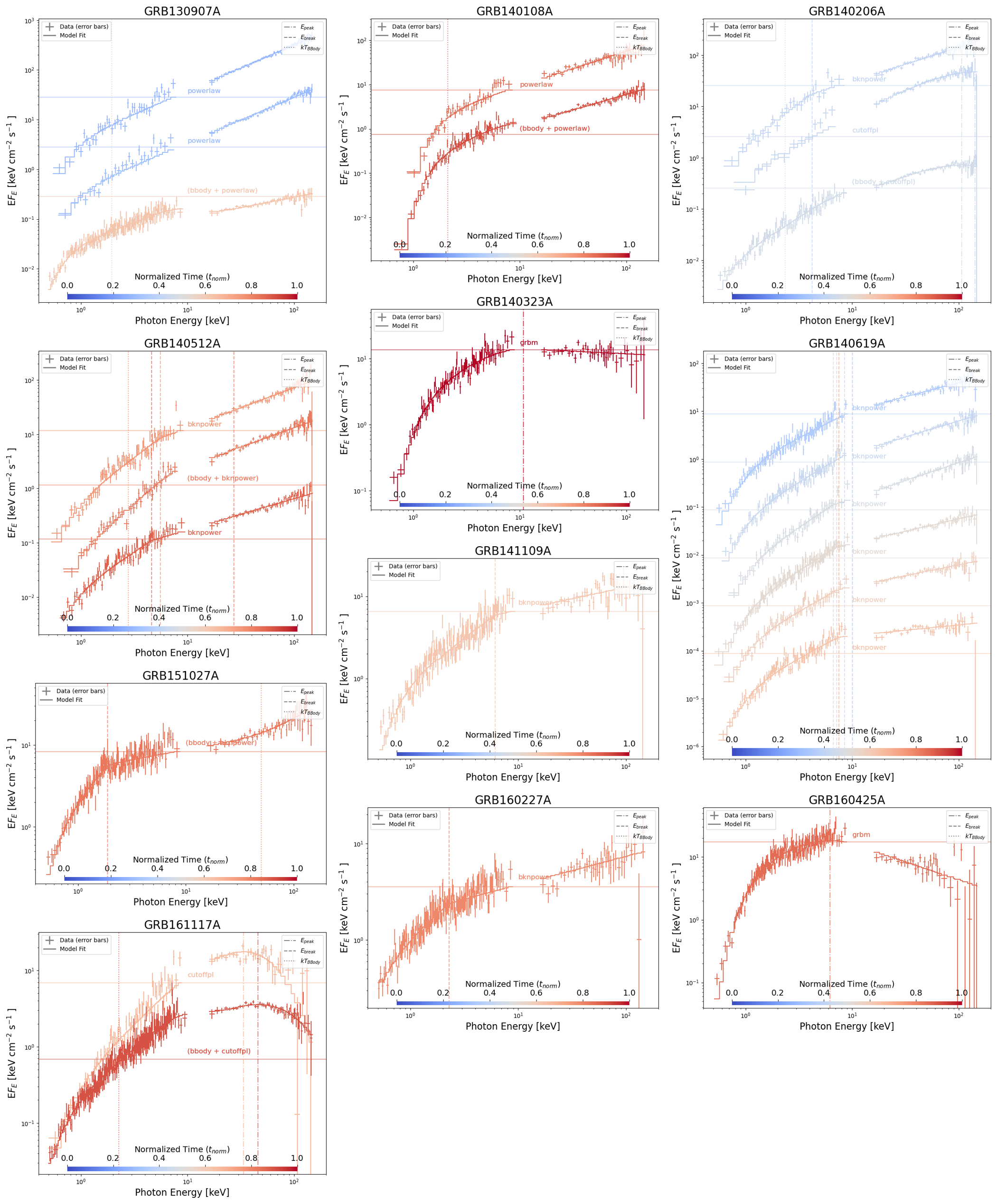}
    \label{fig:placeholder}
\end{figure}
\begin{figure}[h]
    \centering
    \includegraphics[width=1\linewidth]{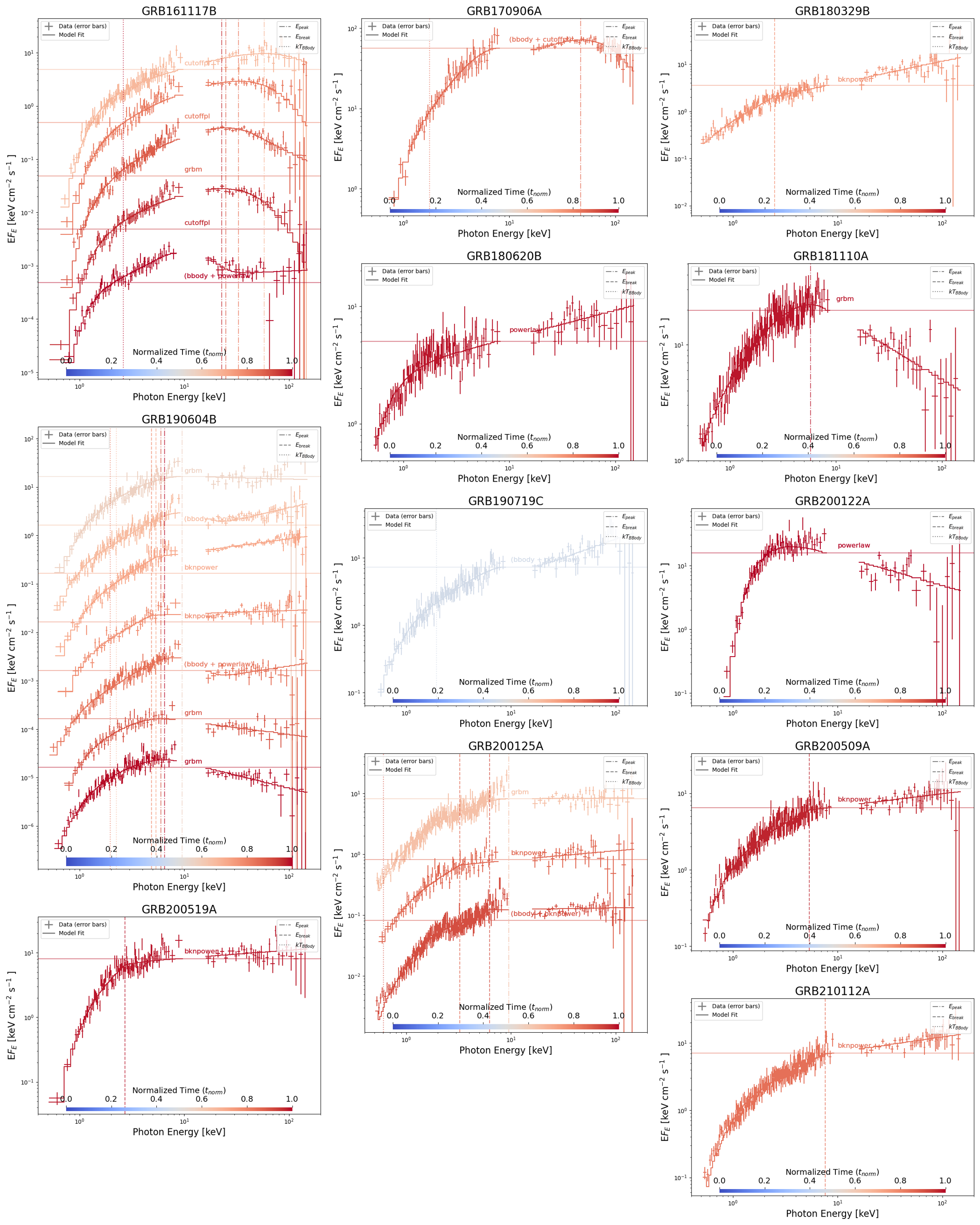}
    \label{fig:placeholder}
\end{figure}
\begin{figure}[h]
    \centering
    \includegraphics[width=1\linewidth]{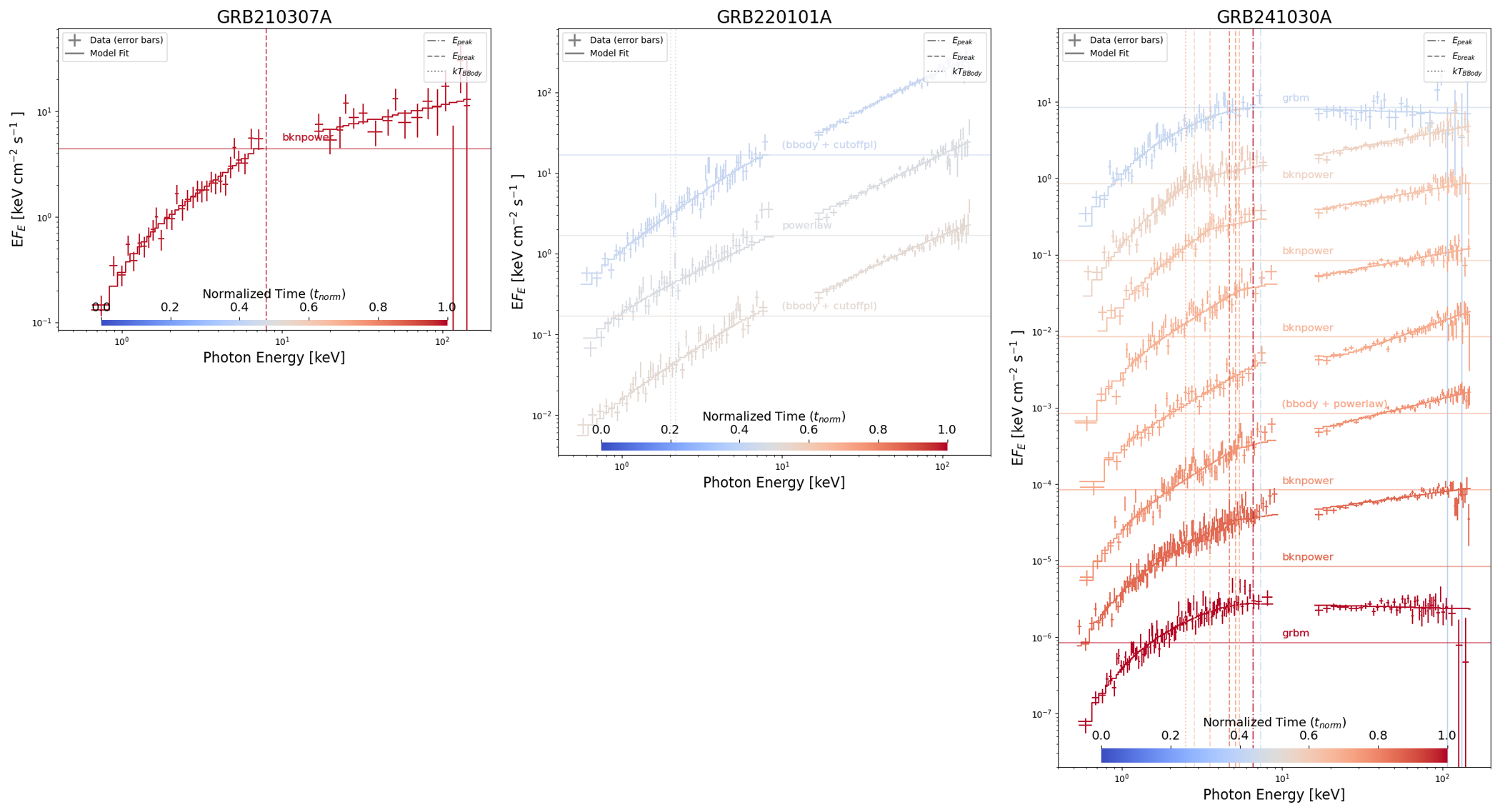}
    \label{fig:placeholder}
\end{figure}

\clearpage
\section{Absorption Parameters} \label{absorption_params}

\begin{table}[h]
\centering
\caption{Absorption parameters for analyzed GRBs}
\label{tab:grb_absorption_params}
\begin{tabular}{cccc|cccc}
\hline
GRB & redshift & galactic $N_H$ & intrinsic $N_H$ & GRB & redshift & galactic $N_H$ & intrinsic $N_H$ \\
 &  & $(10^{22}$ cm$^{-2})$ & $(10^{22}$ cm$^{-2})$ &  &  & $(10^{22}$ cm$^{-2})$ & $(10^{22}$ cm$^{-2})$ \\
\hline
241030A & 1.411 & 0.18 & --- & 140206A & 2.730 & 0.07 & 2 \\
220101A & 4.610 & 0.063 & 0.6 & 140108A & --- & 0.85 & 0.28 \\
210307A & --- & 0.14 & 0.08 & 130907A & 1.238 & 0.01 & 1.1 \\
210112A & --- & 0.0095 & 0.29 & 130609B & --- & 0.015 & 0.16 \\
200519A & --- & 0.33 & 0.13 & 130606A & 5.910 & 0.021 & 2.5 \\
200509A & --- & 0.16 & 0.06 & 130514A & --- & 0.14 & 0.19 \\
200125A & --- & 0.077 & 0.07 & 121217A & --- & 0.52 & --- \\
200122A & --- & 0.015 & 1.8 & 121123A & --- & 0.048 & 0.043 \\
190719C & --- & 0.047 & 0.23 & 110205A & 2.220 & 0.017 & 0.52 \\
190604B & --- & 0.14 & 0.19 & 110119A & --- & 0.087 & 0.08 \\
181110A & 1.505 & 0.077 & 0.02 & 110102A & --- & 0.061 & 0.14 \\
180620B & 1.117 & 0.014 & 0.74 & 100906A & 1.727 & 0.35 & 0.6 \\
180329B & 1.998 & 0.025 & 0.28 & 100814A & --- & 0.018 & --- \\
170906A & --- & 0.14 & 0.33 & 100728A & 1.567 & 0.17 & 2.5 \\
161117B & --- & 0.065 & 0.8 & 100725B & --- & 0.07 & 0.58 \\
161117A & 1.549 & 0.044 & 0.99 & 100619A & --- & 0.022 & 0.68 \\
160425A & 0.555 & 0.077 & 1 & 090715B & 3.000 & 0.014 & 1.3 \\
160227A & 2.380 & 0.041 & 0.4 & 090709A & --- & 0.084 & --- \\
151027A & 0.810 & 0.037 & 0.44 & 081008 & 1.967 & 0.093 & 0.28 \\
141109A & 2.993 & 0.033 & 3.3 & 080928 & 1.690 & 0.072 & 0.36 \\
140619A & --- & 0.017 & 0.27 & 070616 & --- & 0.48 & 0.005 \\
140512A & 0.725 & 0.15 & 0.04 & 061121 & 1.314 & 0.046 & 0.76 \\
140323A & --- & 0.079 & 0.44 & 060814 & 1.923 & 0.026 & 4.3 \\

\hline
\end{tabular}
\end{table}

\clearpage

\section{Spectral Parameters} \label{all-params}

\begin{table}[h]
\centering
\caption{Intervals and Best-fit Parameters for Time-resolved Spectra with $1\sigma$ Errors}
\label{tab:grb_spectral_params}
\begin{tabular}{cccccccccc}
\hline
GRB & Time Bin & $t_{\rm norm}$ & Best Model & $\alpha_1$ & $E_{\rm break}$ & $\alpha_2/\alpha$ & $E_{\rm peak}$ & $\beta$ & $\chi^2$ (dof) \\
 &  (s) &  &  &  & (keV) &  & (keV) &  &  \\
\hline
241030A & [81.93, 96.93] & 0.42 & Band & --- & $5.29^{+1.46}_{-1.17}$ & $-0.62^{+0.16}_{-0.13}$ & $7.32^{+2.18}_{-1.77}$ & $-2.07^{+0.06}_{-0.07}$ & 117.94(114) \\
241030A & [96.93, 118.93] & 0.57 & BPL & $-0.51^{+0.13}_{-0.11}$ & $2.83^{+3.06}_{-2.60}$ & $-1.60^{+0.01}_{-0.02}$ & --- & --- & 149.54(137) \\
241030A & [118.93, 126.93] & 0.63 & BPL & $-0.53^{+0.10}_{-0.10}$ & $3.55^{+3.82}_{-3.30}$ & $-1.63^{+0.02}_{-0.02}$ & --- & --- & 90.41(101) \\
241030A & [126.93, 136.00] & 0.68 & BPL & $-0.77^{+0.06}_{-0.05}$ & $5.42^{+5.87}_{-4.97}$ & $-1.62^{+0.02}_{-0.03}$ & --- & --- & 122.84(114) \\
241030A & [136.00, 141.93] & 0.71 & BBSPL & $-1.26^{+0.02}_{-0.01}$ & --- & $-1.26^{+0.02}_{-0.01}$ & --- & --- & 88.62(93) \\
241030A & [141.93, 157.00] & 0.77 & BPL & $-0.77^{+0.05}_{-0.05}$ & $5.13^{+5.54}_{-4.68}$ & $-1.49^{+0.02}_{-0.02}$ & --- & --- & 143.68(141) \\
241030A & [157.00, 182.00] & 0.86 & BPL & $-0.90^{+0.03}_{-0.04}$ & $4.68^{+5.03}_{-4.34}$ & $-1.72^{+0.02}_{-0.01}$ & --- & --- & 244.80(227) \\
241030A & [182.00, 195.93] & 1.00 & Band & --- & $4.54^{+0.84}_{-0.71}$ & $-0.54^{+0.12}_{-0.11}$ & $6.61^{+1.35}_{-1.16}$ & $-2.05^{+0.03}_{-0.03}$ & 154.20(146) \\
220101A & [110.13, 119.13] & 0.53 & BBCPL & $-0.96^{+0.03}_{-0.01}$ & --- & $-0.96^{+0.03}_{-0.01}$ & $474.45^{+13.69}_{-110.39}$ & --- & 105.89(108) \\
220101A & [101.63, 110.13] & 0.49 & SPL & $-1.06^{+0.01}_{-0.00}$ & --- & $-1.06^{+0.01}_{-0.00}$ & --- & --- & 122.76(108) \\
220101A & [87.13, 101.63] & 0.43 & BBCPL & $-0.85^{+0.02}_{-0.02}$ & --- & $-0.85^{+0.02}_{-0.02}$ & $420.22^{+104.02}_{-70.08}$ & --- & 130.66(119) \\
210307A & [112.48, 120.48] & 0.98 & BPL & $-0.92^{+0.08}_{-0.08}$ & $8.00^{+10.78}_{-6.36}$ & $-1.67^{+0.10}_{-0.15}$ & --- & --- & 68.06(84) \\
210112A & [94.39, 178.39] & 0.84 & BPL & $-1.24^{+0.02}_{-0.02}$ & $7.61^{+8.29}_{-6.82}$ & $-1.78^{+0.03}_{-0.02}$ & --- & --- & 324.56(320) \\
200519A & [66.80, 87.80] & 0.98 & BPL & $-0.73^{+0.14}_{-0.11}$ & $2.72^{+2.94}_{-2.51}$ & $-1.88^{+0.02}_{-0.02}$ & --- & --- & 175.25(156) \\
200509A & [790.00, 848.16] & 0.97 & BPL & $-1.28^{+0.03}_{-0.03}$ & $5.36^{+6.01}_{-4.64}$ & $-1.84^{+0.04}_{-0.03}$ & --- & --- & 294.00(280) \\
200125A & [170.36, 240.00] & 0.64 & Band & --- & $8.62^{+1.37}_{-1.25}$ & $-0.89^{+0.06}_{-0.05}$ & $9.56^{+1.60}_{-1.45}$ & $-1.99^{+0.04}_{-0.04}$ & 401.99(327) \\
200125A & [270.00, 282.98] & 0.86 & BPL & $-0.99^{+0.10}_{-0.08}$ & $3.25^{+3.66}_{-2.82}$ & $-1.85^{+0.04}_{-0.03}$ & --- & --- & 140.87(125) \\
200125A & [282.98, 336.98] & 0.91 & BBBPL & $-0.66^{+0.14}_{-0.11}$ & $6.25^{+6.76}_{-5.92}$ & $-1.97^{+0.03}_{-0.04}$ & --- & --- & 382.28(351) \\
200122A & [188.83, 200.83] & 1.00 & SPL & --- & --- & $-2.47^{+0.02}_{-0.03}$ & --- & $-2.47^{+0.02}_{-0.03}$ & 168.82(132) \\
190719C & [76.75, 102.84] & 0.46 & BBSPL & $-1.45^{+0.02}_{-0.01}$ & --- & $-1.45^{+0.02}_{-0.01}$ & --- & --- & 211.42(186) \\
190604B & [120.56, 130.56] & 0.57 & Band & --- & $6.69^{+1.17}_{-1.03}$ & $-0.58^{+0.10}_{-0.09}$ & $9.52^{+1.78}_{-1.58}$ & $-2.06^{+0.06}_{-0.07}$ & 167.73(155) \\
190604B & [130.56, 148.00] & 0.65 & BBSPL & $-1.50^{+0.01}_{-0.02}$ & --- & $-1.50^{+0.01}_{-0.02}$ & --- & --- & 250.30(184) \\
190604B & [148.00, 167.00] & 0.72 & BPL & $-0.72^{+0.06}_{-0.06}$ & $5.36^{+5.84}_{-4.97}$ & $-1.71^{+0.02}_{-0.02}$ & --- & --- & 138.03(133) \\
190604B & [167.00, 178.00] & 0.78 & BPL & $-0.91^{+0.09}_{-0.09}$ & $4.84^{+5.52}_{-4.40}$ & $-1.93^{+0.04}_{-0.05}$ & --- & --- & 116.39(93) \\
190604B & [178.00, 190.68] & 0.84 & BBSPL & $-1.66^{+0.02}_{-0.02}$ & --- & $-1.66^{+0.02}_{-0.02}$ & --- & --- & 191.57(166) \\
190604B & [190.68, 208.00] & 0.90 & Band & --- & $4.01^{+1.10}_{-0.95}$ & $-0.51^{+0.23}_{-0.17}$ & $5.96^{+1.87}_{-1.57}$ & $-2.28^{+0.06}_{-0.07}$ & 103.81(105) \\
190604B & [208.00, 238.68] & 0.99 & Band & --- & $4.48^{+0.60}_{-0.53}$ & $-0.56^{+0.10}_{-0.10}$ & $6.47^{+0.98}_{-0.88}$ & $-2.53^{+0.06}_{-0.06}$ & 191.98(169) \\
181110A & [67.79, 105.28] & 0.98 & Band & --- & $3.95^{+0.47}_{-0.43}$ & $-0.55^{+0.08}_{-0.07}$ & $5.73^{+0.74}_{-0.68}$ & $-2.55^{+0.05}_{-0.06}$ & 275.58(260) \\
180620B & [92.20, 116.20] & 0.98 & SPL & $-1.76^{+0.01}_{-0.02}$ & --- & $-1.76^{+0.01}_{-0.02}$ & --- & --- & 182.69(174) \\
180329B & [137.83, 173.83] & 0.77 & BPL & $-0.80^{+0.09}_{-0.09}$ & $2.50^{+2.88}_{-2.24}$ & $-1.53^{+0.02}_{-0.03}$ & --- & --- & 214.54(180) \\
170906A & [75.15, 90.15] & 0.85 & BBCPL & $-1.08^{+0.05}_{-0.04}$ & --- & $-1.08^{+0.05}_{-0.04}$ & $46.96^{+3.95}_{-3.51}$ & --- & 87.19(112) \\
161117B & [129.60, 143.60] & 0.88 & Band & --- & $21.00^{+2.00}_{-1.85}$ & $-0.82^{+0.01}_{-0.01}$ & $24.85^{+2.39}_{-2.22}$ & $-3.00^{+0.13}_{-0.16}$ & 184.70(151) \\
161117B & [80.71, 117.60] & 0.66 & CPL & $-1.33^{+0.03}_{-0.03}$ & --- & $-1.33^{+0.03}_{-0.03}$ & $57.69^{+10.89}_{-8.43}$ & --- & 237.58(182) \\
161117B & [117.60, 129.60] & 0.83 & CPL & $-1.00^{+0.03}_{-0.04}$ & --- & $-1.00^{+0.03}_{-0.04}$ & $33.09^{+2.65}_{-2.58}$ & --- & 132.91(120) \\
161117B & [143.60, 152.60] & 0.97 & CPL & $-1.04^{+0.04}_{-0.04}$ & --- & $-1.04^{+0.04}_{-0.04}$ & $22.72^{+2.10}_{-1.92}$ & --- & 135.76(122) \\
161117B & [152.60, 161.60] & 1.02 & BBSPL & $-1.92^{+0.03}_{-0.03}$ & --- & $-1.92^{+0.03}_{-0.03}$ & --- & --- & 112.87(117) \\
161117A & [67.14, 92.14] & 0.67 & CPL & $-0.50^{+0.03}_{-0.04}$ & --- & $-0.50^{+0.03}_{-0.04}$ & $33.60^{+2.03}_{-2.00}$ & --- & 174.34(128) \\
161117A & [92.14, 174.64] & 0.90 & BBCPL & $-1.09^{+0.02}_{-0.01}$ & --- & $-1.09^{+0.02}_{-0.01}$ & $45.87^{+1.97}_{-1.66}$ & --- & 339.01(353) \\
160425A & [257.08, 295.98] & 0.85 & Band & --- & $6.05^{+0.71}_{-0.65}$ & $-0.97^{+0.06}_{-0.06}$ & $6.21^{+0.83}_{-0.76}$ & $-2.57^{+0.05}_{-0.06}$ & 296.70(266) \\
160227A & [182.20, 236.50] & 0.79 & BPL & $-1.08^{+0.05}_{-0.06}$ & $2.29^{+2.48}_{-2.11}$ & $-1.71^{+0.02}_{-0.01}$ & --- & --- & 263.63(254) \\

\hline
\end{tabular}
\end{table}

\begin{table}[h]
\centering
\caption{Intervals and Best-fit Parameters for Time-resolved Spectra with $1\sigma$ Errors (continued)}
\begin{tabular}{cccccccccc}
\hline
GRB & Time Bin & $t_{\rm norm}$ & Best Model & $\alpha_1$ & $E_{\rm break}$ & $\alpha_2/\alpha$ & $E_{\rm peak}$ & $\beta$ & $\chi^2$ (dof) \\
 &  (s) &  &  &  & (keV) &  & (keV) &  &  \\
\hline
151027A & [93.27, 164.01] & 0.83 & BBBPL & $-0.72^{+0.08}_{-0.09}$ & $1.87^{+1.96}_{-1.81}$ & $-1.77^{+0.01}_{-0.02}$ & --- & --- & 229.01(227) \\
141109A & [132.98, 178.50] & 0.62 & BPL & $-1.05^{+0.02}_{-0.03}$ & $6.14^{+6.63}_{-5.65}$ & $-1.72^{+0.02}_{-0.03}$ & --- & --- & 345.25(241) \\
140619A & [169.50, 188.50] & 0.43 & BPL & $-0.72^{+0.04}_{-0.04}$ & $8.49^{+9.50}_{-7.72}$ & $-1.33^{+0.03}_{-0.03}$ & --- & --- & 163.03(154) \\
140619A & [145.37, 169.50] & 0.36 & BPL & $-0.90^{+0.04}_{-0.04}$ & $10.08^{+11.78}_{-8.70}$ & $-1.42^{+0.04}_{-0.03}$ & --- & --- & 166.63(165) \\
140619A & [197.50, 215.50] & 0.52 & BPL & $-0.64^{+0.03}_{-0.04}$ & $7.16^{+7.66}_{-6.72}$ & $-1.49^{+0.03}_{-0.02}$ & --- & --- & 176.64(184) \\
140619A & [225.50, 241.37] & 0.63 & BPL & $-0.83^{+0.04}_{-0.04}$ & $7.46^{+7.99}_{-6.90}$ & $-1.78^{+0.04}_{-0.03}$ & --- & --- & 198.93(154) \\
140619A & [188.50, 197.50] & 0.48 & BPL & $-0.57^{+0.07}_{-0.06}$ & $6.64^{+7.46}_{-5.99}$ & $-1.26^{+0.03}_{-0.02}$ & --- & --- & 112.43(109) \\
140619A & [215.50, 225.50] & 0.58 & BPL & $-0.79^{+0.06}_{-0.05}$ & $7.52^{+8.34}_{-6.65}$ & $-1.55^{+0.03}_{-0.03}$ & --- & --- & 99.12(115) \\
140512A & [124.28, 140.28] & 0.87 & BPL & $-0.74^{+0.05}_{-0.05}$ & $4.60^{+4.96}_{-4.26}$ & $-1.42^{+0.01}_{-0.02}$ & --- & --- & 135.24(138) \\
140512A & [102.15, 118.28] & 0.76 & BPL & $-0.56^{+0.05}_{-0.05}$ & $5.55^{+6.10}_{-5.11}$ & $-1.26^{+0.02}_{-0.02}$ & --- & --- & 144.92(123) \\
140512A & [118.28, 124.28] & 0.82 & BBBPL & $-0.71^{+0.05}_{-0.05}$ & $27.33^{+31.32}_{-23.43}$ & $-1.29^{+0.04}_{-0.03}$ & --- & --- & 80.30(87) \\
140323A & [101.22, 119.79] & 1.00 & Band & --- & $9.37^{+37.42}_{-1.54}$ & $-0.85^{+0.09}_{-0.06}$ & $10.80^{+43.12}_{-1.86}$ & $-2.08^{+0.05}_{-0.05}$ & 162.02(145) \\
140206A & [59.23, 96.72] & 0.46 & BBCPL & $-0.98^{+0.02}_{-0.02}$ & --- & $-0.98^{+0.02}_{-0.02}$ & $107.40^{+7.62}_{-6.82}$ & --- & 136.49(145) \\
140206A & [57.23, 59.23] & 0.43 & CPL & $-0.85^{+0.03}_{-0.03}$ & --- & $-0.85^{+0.03}_{-0.03}$ & $142.65^{+14.88}_{-12.83}$ & --- & 59.34(67) \\
140206A & [50.23, 57.23] & 0.41 & BPL & $-0.55^{+0.14}_{-0.11}$ & $4.20^{+4.78}_{-3.63}$ & $-1.34^{+0.02}_{-0.02}$ & --- & --- & 71.41(79) \\
140108A & [82.99, 103.48] & 0.90 & BBSPL & $-1.30^{+0.02}_{-0.01}$ & --- & $-1.30^{+0.02}_{-0.01}$ & --- & --- & 135.67(136) \\
140108A & [74.99, 82.99] & 0.84 & SPL & $-1.21^{+0.01}_{-0.02}$ & --- & $-1.21^{+0.01}_{-0.02}$ & --- & --- & 93.00(93) \\
130907A & [74.66, 82.00] & 0.28 & SPL & $-1.08^{+0.01}_{-0.01}$ & --- & $-1.08^{+0.01}_{-0.01}$ & --- & --- & 86.88(94) \\
130907A & [82.00, 87.66] & 0.29 & SPL & $-1.09^{+0.02}_{-0.01}$ & --- & $-1.09^{+0.02}_{-0.01}$ & --- & --- & 93.73(84) \\
130907A & [200.00, 270.00] & 0.62 & BBSPL & $-1.58^{+0.01}_{-0.01}$ & --- & $-1.58^{+0.01}_{-0.01}$ & --- & --- & 234.43(257) \\
130609B & [162.29, 232.43] & 0.88 & BBSPL & $-1.75^{+0.01}_{-0.01}$ & --- & $-1.75^{+0.01}_{-0.01}$ & --- & --- & 295.52(284) \\
130606A & [140.00, 165.00] & 0.35 & SPL & $-1.10^{+0.01}_{-0.01}$ & --- & $-1.10^{+0.01}_{-0.01}$ & --- & --- & 120.85(115) \\
130514A & [95.12, 175.88] & 0.61 & Band & --- & $11.53^{+1.33}_{-2.44}$ & $-1.09^{+0.06}_{-0.02}$ & $10.52^{+1.42}_{-2.25}$ & $-2.12^{+0.04}_{-0.03}$ & 245.39(213) \\
121217A & [710.00, 800.00] & 0.90 & BPL & $-1.26^{+0.03}_{-0.02}$ & $7.03^{+7.82}_{-6.38}$ & $-1.73^{+0.02}_{-0.03}$ & --- & --- & 257.08(280) \\
121123A & [196.00, 214.00] & 0.53 & CPL & $-0.77^{+0.02}_{-0.02}$ & --- & $-0.77^{+0.02}_{-0.02}$ & $127.35^{+14.37}_{-12.00}$ & --- & 97.29(114) \\
121123A & [270.00, 300.00] & 0.76 & BBCPL & $-1.09^{+0.03}_{-0.04}$ & --- & $-1.09^{+0.03}_{-0.04}$ & $45.53^{+5.08}_{-4.64}$ & --- & 215.93(211) \\
121123A & [252.00, 270.00] & 0.69 & BBCPL & $-0.92^{+0.03}_{-0.03}$ & --- & $-0.92^{+0.03}_{-0.03}$ & $46.76^{+3.85}_{-3.45}$ & --- & 137.11(169) \\
121123A & [239.00, 252.00] & 0.65 & BBCPL & $-0.73^{+0.04}_{-0.03}$ & --- & $-0.73^{+0.04}_{-0.03}$ & $53.96^{+3.90}_{-3.46}$ & --- & 154.55(158) \\
121123A & [214.00, 239.00] & 0.60 & CPL & $-0.78^{+0.01}_{-0.02}$ & --- & $-0.78^{+0.01}_{-0.02}$ & $93.79^{+5.22}_{-4.95}$ & --- & 180.50(166) \\
110205A & [226.00, 240.00] & 0.59 & BPL & $-0.71^{+0.07}_{-0.06}$ & $5.38^{+6.22}_{-4.65}$ & $-1.81^{+0.05}_{-0.05}$ & --- & --- & 93.32(114) \\
110205A & [194.00, 226.00] & 0.53 & BPL & $-0.68^{+0.05}_{-0.03}$ & $5.56^{+5.97}_{-5.00}$ & $-1.64^{+0.03}_{-0.02}$ & --- & --- & 174.78(191) \\
110205A & [159.20, 194.00] & 0.42 & BPL & $-0.60^{+0.03}_{-0.03}$ & $5.64^{+5.88}_{-5.34}$ & $-1.85^{+0.02}_{-0.02}$ & --- & --- & 199.19(197) \\
110119A & [163.50, 186.50] & 0.82 & CPL & $-1.07^{+0.04}_{-0.04}$ & --- & $-1.07^{+0.04}_{-0.04}$ & $74.85^{+19.73}_{-13.92}$ & --- & 104.85(123) \\
110119A & [186.50, 225.50] & 0.97 & BPL & $-0.99^{+0.03}_{-0.03}$ & $5.89^{+6.43}_{-5.23}$ & $-1.73^{+0.04}_{-0.03}$ & --- & --- & 212.27(181) \\
110102A & [256.50, 306.50] & 0.98 & BPL & $-1.15^{+0.03}_{-0.03}$ & $4.56^{+4.84}_{-4.20}$ & $-1.84^{+0.02}_{-0.02}$ & --- & --- & 333.75(312) \\
110102A & [220.50, 242.50] & 0.83 & BPL & $-1.25^{+0.04}_{-0.04}$ & $5.96^{+6.85}_{-5.36}$ & $-1.91^{+0.04}_{-0.05}$ & --- & --- & 168.88(156) \\
110102A & [242.50, 256.50] & 0.91 & BPL & $-1.01^{+0.07}_{-0.06}$ & $3.78^{+4.13}_{-3.48}$ & $-1.85^{+0.03}_{-0.03}$ & --- & --- & 129.23(129) \\
110102A & [194.50, 220.50] & 0.76 & BPL & $-0.68^{+0.04}_{-0.03}$ & $6.67^{+7.16}_{-6.20}$ & $-1.38^{+0.02}_{-0.02}$ & --- & --- & 216.28(202) \\
100906A & [113.00, 135.00] & 1.03 & Band & --- & $3.25^{+0.34}_{-0.32}$ & $-0.35^{+0.10}_{-0.09}$ & $5.37^{+0.66}_{-0.61}$ & $-3.13^{+0.08}_{-0.09}$ & 216.74(215) \\
100906A & [102.50, 113.00] & 0.92 & Band & --- & $4.96^{+0.71}_{-0.66}$ & $-0.36^{+0.13}_{-0.12}$ & $8.15^{+1.34}_{-1.23}$ & $-2.45^{+0.05}_{-0.05}$ & 111.15(119) \\
100906A & [84.10, 102.50] & 0.85 & Band & --- & $6.03^{+1.40}_{-1.18}$ & $-0.78^{+0.14}_{-0.12}$ & $7.34^{+1.90}_{-1.62}$ & $-2.30^{+0.06}_{-0.07}$ & 125.64(116) \\
100814A & [94.00, 172.00] & 0.77 & BPL & $-1.10^{+0.02}_{-0.02}$ & $6.27^{+6.83}_{-5.83}$ & $-1.86^{+0.03}_{-0.03}$ & --- & --- & 283.86(287) \\
100728A & [118.00, 139.00] & 0.82 & CPL & $-1.22^{+0.01}_{-0.02}$ & --- & $-1.22^{+0.01}_{-0.02}$ & $206.72^{+41.64}_{-30.38}$ & --- & 185.36(198) \\
100728A & [81.00, 94.00] & 0.63 & BPL & $-1.08^{+0.06}_{-0.05}$ & $6.19^{+7.05}_{-4.97}$ & $-1.52^{+0.03}_{-0.03}$ & --- & --- & 147.86(133) \\
100728A & [94.00, 105.00] & 0.69 & SPL & $-1.42^{+0.01}_{-0.01}$ & --- & $-1.42^{+0.01}_{-0.01}$ & --- & --- & 123.30(124) \\
100728A & [105.00, 118.00] & 0.76 & SPL & $-1.28^{+0.01}_{-0.00}$ & --- & $-1.28^{+0.01}_{-0.00}$ & --- & --- & 118.50(132) \\
100728A & [139.00, 159.00] & 0.94 & CPL & $-1.39^{+0.03}_{-0.04}$ & --- & $-1.39^{+0.03}_{-0.04}$ & $42.98^{+9.40}_{-7.27}$ & --- & 135.95(137) \\
100725B & [143.00, 170.39] & 0.92 & BBBPL & $-1.31^{+0.12}_{-0.12}$ & $2.29^{+2.61}_{-2.07}$ & $-1.95^{+0.03}_{-0.03}$ & --- & --- & 236.51(211) \\
100725B & [207.00, 219.78] & 1.33 & CPL & $-1.37^{+0.08}_{-0.07}$ & --- & $-1.37^{+0.08}_{-0.07}$ & $6.85^{+1.17}_{-1.00}$ & --- & 108.57(118) \\
100725B & [85.00, 123.00] & 0.72 & SPL & $-1.57^{+0.01}_{-0.01}$ & --- & $-1.57^{+0.01}_{-0.01}$ & --- & --- & 220.43(214) \\

\hline
\end{tabular}
\end{table}

\begin{table}[h]
\centering
\caption{Intervals and Best-fit Parameters for Time-resolved Spectra with $1\sigma$ Errors (continued)}
\begin{tabular}{cccccccccc}
\hline
GRB & Time Bin & $t_{\rm norm}$ & Best Model & $\alpha_1$ & $E_{\rm break}$ & $\alpha_2/\alpha$ & $E_{\rm peak}$ & $\beta$ & $\chi^2$ (dof) \\
 &  (s) &  &  &  & (keV) &  & (keV) &  &  \\
\hline
100906A & [84.10, 102.50] & 0.85 & Band & --- & $6.03^{+1.40}_{-1.18}$ & $-0.78^{+0.14}_{-0.12}$ & $7.34^{+1.90}_{-1.62}$ & $-2.30^{+0.06}_{-0.07}$ & 125.64(116) \\
100814A & [94.00, 172.00] & 0.77 & BPL & $-1.10^{+0.02}_{-0.02}$ & $6.27^{+6.83}_{-5.83}$ & $-1.86^{+0.03}_{-0.03}$ & --- & --- & 283.86(287) \\
100728A & [118.00, 139.00] & 0.82 & CPL & $-1.22^{+0.01}_{-0.02}$ & --- & $-1.22^{+0.01}_{-0.02}$ & $206.72^{+41.64}_{-30.38}$ & --- & 185.36(198) \\
100728A & [81.00, 94.00] & 0.63 & BPL & $-1.08^{+0.06}_{-0.05}$ & $6.19^{+7.05}_{-4.97}$ & $-1.52^{+0.03}_{-0.03}$ & --- & --- & 147.86(133) \\
100728A & [94.00, 105.00] & 0.69 & SPL & $-1.42^{+0.01}_{-0.01}$ & --- & $-1.42^{+0.01}_{-0.01}$ & --- & --- & 123.30(124) \\
100728A & [105.00, 118.00] & 0.76 & SPL & $-1.28^{+0.01}_{-0.00}$ & --- & $-1.28^{+0.01}_{-0.00}$ & --- & --- & 118.50(132) \\
100728A & [139.00, 159.00] & 0.94 & CPL & $-1.39^{+0.03}_{-0.04}$ & --- & $-1.39^{+0.03}_{-0.04}$ & $42.98^{+9.40}_{-7.27}$ & --- & 135.95(137) \\
100725B & [143.00, 170.39] & 0.92 & BBBPL & $-1.31^{+0.12}_{-0.12}$ & $2.29^{+2.61}_{-2.07}$ & $-1.95^{+0.03}_{-0.03}$ & --- & --- & 236.51(211) \\
100725B & [207.00, 219.78] & 1.33 & CPL & $-1.37^{+0.08}_{-0.07}$ & --- & $-1.37^{+0.08}_{-0.07}$ & $6.85^{+1.17}_{-1.00}$ & --- & 108.57(118) \\
100725B & [85.00, 123.00] & 0.72 & SPL & $-1.57^{+0.01}_{-0.01}$ & --- & $-1.57^{+0.01}_{-0.01}$ & --- & --- & 220.43(214) \\
100725B & [123.00, 130.00] & 0.78 & BPL & $-0.87^{+0.06}_{-0.07}$ & $9.81^{+11.14}_{-8.76}$ & $-1.72^{+0.04}_{-0.05}$ & --- & --- & 104.72(107) \\
100725B & [130.00, 136.60] & 0.82 & CPL & $-1.12^{+0.03}_{-0.03}$ & --- & $-1.12^{+0.03}_{-0.03}$ & $66.81^{+8.06}_{-6.85}$ & --- & 134.10(111) \\
100725B & [136.60, 143.00] & 0.86 & CPL & $-1.22^{+0.06}_{-0.05}$ & --- & $-1.22^{+0.06}_{-0.05}$ & $26.48^{+4.55}_{-3.74}$ & --- & 124.77(108) \\
100619A & [80.35, 95.00] & 0.89 & BPL & $-1.04^{+0.06}_{-0.05}$ & $6.73^{+7.29}_{-6.27}$ & $-1.98^{+0.03}_{-0.02}$ & --- & --- & 118.59(124) \\
100619A & [95.00, 107.35] & 1.01 & SPL & $-1.98^{+0.02}_{-0.02}$ & --- & $-1.98^{+0.02}_{-0.02}$ & --- & --- & 89.27(85) \\
090715B & [52.46, 100.95] & 0.28 & BPL & $-0.99^{+0.02}_{-0.03}$ & $5.68^{+6.38}_{-5.30}$ & $-1.75^{+0.02}_{-0.04}$ & --- & --- & 185.07(200) \\
090709A & [74.10, 83.00] & 0.86 & BPL & $-0.60^{+0.07}_{-0.09}$ & $5.15^{+6.14}_{-4.60}$ & $-1.54^{+0.03}_{-0.06}$ & --- & --- & 99.51(100) \\
090709A & [83.00, 110.00] & 0.97 & BPL & $-0.63^{+0.09}_{-0.06}$ & $3.77^{+4.12}_{-3.24}$ & $-1.47^{+0.02}_{-0.02}$ & --- & --- & 136.81(148) \\
081008 & [90.96, 144.96] & 0.86 & BPL & $-1.09^{+0.02}_{-0.02}$ & $5.29^{+5.63}_{-4.79}$ & $-1.95^{+0.04}_{-0.03}$ & --- & --- & 364.07(318) \\
080928 & [194.50, 230.00] & 0.61 & BPL & $-1.02^{+0.03}_{-0.03}$ & $4.65^{+4.89}_{-4.38}$ & $-1.99^{+0.02}_{-0.03}$ & --- & --- & 206.04(218) \\
070616 & [473.00, 530.00] & 0.90 & BPL & $-1.26^{+0.03}_{-0.03}$ & $7.08^{+8.46}_{-6.41}$ & $-2.00^{+0.05}_{-0.07}$ & --- & --- & 202.39(231) \\
070616 & [185.00, 223.00] & 0.27 & BPL & $-1.01^{+0.03}_{-0.04}$ & $6.34^{+7.20}_{-5.82}$ & $-1.52^{+0.02}_{-0.03}$ & --- & --- & 180.30(190) \\
070616 & [398.00, 473.00] & 0.75 & BPL & $-1.12^{+0.04}_{-0.03}$ & $5.21^{+5.62}_{-4.66}$ & $-1.73^{+0.02}_{-0.02}$ & --- & --- & 287.45(238) \\
070616 & [137.00, 185.00] & 0.15 & BPL & $-1.05^{+0.03}_{-0.03}$ & $8.41^{+9.77}_{-7.37}$ & $-1.46^{+0.03}_{-0.03}$ & --- & --- & 156.16(185) \\
070616 & [223.00, 285.00] & 0.39 & BPL & $-0.95^{+0.02}_{-0.03}$ & $7.17^{+7.65}_{-6.60}$ & $-1.52^{+0.02}_{-0.02}$ & --- & --- & 267.00(271) \\
070616 & [285.00, 330.00] & 0.50 & BPL & $-0.97^{+0.04}_{-0.03}$ & $6.63^{+7.24}_{-6.01}$ & $-1.49^{+0.02}_{-0.03}$ & --- & --- & 129.02(181) \\
070616 & [330.00, 398.00] & 0.63 & BPL & $-1.09^{+0.03}_{-0.03}$ & $5.28^{+5.74}_{-4.78}$ & $-1.66^{+0.02}_{-0.02}$ & --- & --- & 327.10(298) \\
061121 & [66.00, 70.00] & 0.30 & BPL & $-0.67^{+0.07}_{-0.07}$ & $7.33^{+8.47}_{-6.50}$ & $-1.34^{+0.02}_{-0.02}$ & --- & --- & 85.21(83) \\
061121 & [70.00, 78.00] & 0.48 & BPL & $-0.53^{+0.06}_{-0.06}$ & $5.25^{+5.66}_{-4.85}$ & $-1.34^{+0.01}_{-0.01}$ & --- & --- & 82.51(124) \\
060814 & [115.00, 152.08] & 0.77 & BBSPL & $-1.63^{+0.01}_{-0.01}$ & --- & $-1.63^{+0.01}_{-0.01}$ & --- & --- & 202.90(204) \\

\hline
\end{tabular}
\end{table}

\clearpage
\bibliography{main}{}
\bibliographystyle{aasjournal}



\end{document}